\documentclass[%english,
a4paper,11pt]{amsart}

\usepackage[figuresright]{rotating}
\usepackage{amsmath,amssymb}

\usepackage[dvipsnames]{xcolor}
\usepackage{tikz,tkz-tab}
\usepackage{url}

\newtheorem{lemma}{Lemma}
\newtheorem{definition}{Definition}
\newtheorem{theorem}{Theorem}
\newtheorem{prop}{Proposition}
\newtheorem{coro}{Corollary}

 \def\ZZ{{\mathbb{Z}}}
 \def\RR{{\mathbb{R}}}

  \def\JJ{{\mathcal{J}}}
   \def\SS{{\mathcal{S}}}
  
 \def\PP{{\mathcal{P}}}
\def\infty{\rotatebox{90}{8}}

\date{%version du \today
}

\begin{document}

 \title[Polynomial algorithms for p-dispersion problems in a 2D PF]{Polynomial algorithms for p-dispersion problems in a planar Pareto Front}
% \titlerunning{Polynomial algorithms for $p$-dispersion problems in a 2D PF}
% 
%  \author{Nicolas Dupin%\inst{1}\orcidID{0000-0003-3775-5629} 
%  }
\author{Nicolas Dupin $^{1,*}$ }
\thanks{$^1$ Univ Angers, LERIA, SFR MATHSTIC, F-49000 Angers, France}%Laboratoire Interdiscplinaire des Sciences du Numérique, Université Paris-Saclay, Gif-sur-Yvette, France}
\thanks{$^*$ Corresponding author: nicolas.dupin@univ-angers.fr}

\date{%First submission Dec, 2019 last revision 
version  \today
}

\maketitle

{\noindent \color{Blue} Thi artile should be cited as:}

{\noindent  \color{Blue} N. Dupin,
Polynomial algorithms for p-dispersion problems in a planar Pareto Front,
RAIRO-Oper. Res. 57 (2023) 857–880.
\url{ 	https://doi.org/10.1051/ro/2023034 }
}

\begin{abstract}
In this paper, $p$-dispersion problems  are studied  to select  $p\geqslant 2$ representative points from a large 2D Pareto Front (PF), solution of  bi-objective optimization. % problems,
Four standard  $p$-dispersion variants  are considered.
 A novel variant,  Max-Sum-Neighbor $p$-dispersion, is introduced for the specific case of a 2D PF.
Firstly, $2$-dispersion and  $3$-dispersion problems are proven solvable in $O(n)$  time in a 2D PF.
Secondly, dynamic programming algorithms are designed for three  $p$-dispersion variants, proving polynomial complexities in a 2D PF.
Max-min $p$-dispersion is solvable in $O(pn\log n)$  time  and $O(n)$ memory space.
 Max-Sum-Neighbor $p$-dispersion is proven solvable in $O(pn^2)$  time  and $O(n)$  space.
 Max-Sum-min $p$-dispersion is solvable in $O(pn^3)$  time  and $O(pn^2)$  space. These complexity results hold also in 1D, proving for the first time that  Max-Sum-min $p$-dispersion is polynomial in 1D.
Furthermore,    properties of these algorithms are discussed  for an efficient implementation
and for practical applications. % inside bi-objective  meta-heuristics.
 \end{abstract}
% \noindent{
% \textbf{Keywords} : Optimization ; Algorithms; %, Operational Research, Computational Geometry ;
% Dynamic programming ;
%  $p$-dispersion problems  ;  complexity  ; bi-objective optimization ;  Pareto front ; Skyline operator}

\section{Introduction}

In real-world applications,  optimization problems may be driven by several conflicting objectives. 
 Designing  network or  (system of) system architectures, financial costs must be traded off with quality of service or robustness  \cite{denstad2019multi,peugeot2017mbse}.
% In finance, risk aversion of decision makers induces a trade-off between expected outcomes and uncertainty \cite{masri2017multiple,xidonas2020robust}.
 Dealing with complex maintenance planning problems,  stability and  robustness of the planning matter as well as financial costs \cite{dupin2020matheuristics}.
%Generally,  bi-objective optimization allows to consider robust optimization formulations \cite{kouvelis2006algorithm,bahri2020robustness}.
Multi-objective optimization (MOO) supports such % trade-off
decision making.
% 
%This paper is motivated by real-world applications of multi-objective optimization (MOO). % \cite{dupin2015modelisation,peugeot2017mbse}.
  Many  efficient (i.e. best compromise) solutions of MOO problems may exist with Pareto dominance \cite{ehrgott2003multiobjective}.
 A Pareto Front (PF) denotes the projection of efficient solutions in the  objective space. % space
 This work aims  to select $p$ solutions  from $n\gg p$ non dominated solutions, while
 maximizing the diversity of these $p$ solutions in the objective space.
Firstly, such problem occurs when selecting alternatives for decision makers. %,  and  in such applications $p$ is small.
Secondly, 
MOO approaches  use such operators to represent large PF  \cite{bazgan2017discrete,sayin2000measuring},
and MOO meta-heuristics  archive diversified solutions during the heuristic search  \cite{schuetze2019archivers,talbi2009metaheuristics,zio2011clustering}.
Thirdly,  a similar problem occurs  in a Skyline for databases
\cite{borzsony2001skyline,cabello2020faster,lin2007selecting}.
When  selecting alternatives for decision makers, $p$ is small, $p \leqslant 5$ is realistic.
Otherwise, $p$ is larger, having $p=100$ or $p=1000$ is realistic.

The hypervolume measure is often used in such a context  \cite{auger2009investigating,falcon2020indicator,guerreiro2020hypervolume}.
Covering and clustering algorithms %selecting firstly $p$ clusters and then returning the most central point of the cluster,
 \cite{dupin2019medoids,dupin2021unified,zio2011clustering}
are also used  to select points in a PF.
In this paper, we consider (discrete) $p$-dispersion problems,   to select $p$ points and while maximizing dispersion measures among selected points \cite{erkut1990discrete}.
Although $p$-dispersion is mentioned to  have relevant applications for MOO \cite{erkut1991comparison,ravi1994heuristic}, no specific studies concerned the $p$-dispersion in a PF to the best of our knowledge.
Four variants of discrete $p$-dispersion problems are defined   \cite{erkut1991comparison}.
Max-min and Max-Sum $p$-dispersion problems,  the most studied variants,  are  $\mathcal{N}\mathcal{P}$-hard \cite{erkut1990discrete,hansen1995dispersing}.
Max-min $p$-dispersion maximizes the minimal distance between each pair of selected points.
Max-Sum $p$-dispersion maximizes the total sum of the distances between selected points. % are considered instead as dispersion criterion.
Max-Sum-min $p$-dispersion  variant maximizes
the sum of the distances between each selected points to the closest different selected points.
Max-min-Sum  $p$-dispersion  variant maximizes
the minimal sum of distances between each selected point to the other selected points.
This paper studies  $p$-dispersion variants in the case of a two-dimensional (2D) PF, which is an extension of one-dimensional (1D) cases.
A novel   variant of  Max-Sum-min $p$-dispersion, denoted Max-Sum-Neighbor $p$-dispersion, % variant of the Max-Sum-min $p$-dispersion problem,
is specifically introduced for 2D PFs.
 For these five $p$-dispersion problems,  the cases $p=2$  and $p=3$ are proven to  be solvable in $O(n)$ time and the cases $p=4$  and $p=5$ are   solvable respectively in $O(n^2)$ and $O(n^3)$ time.
 Generally, the Max-min, Max-Sum-min and Max-Sum-Neighbor $p$-dispersion problems are proven to be solvable in polynomial time in a 2D PF with dynamic programming (DP) algorithms.
For Max-Sum-min $p$-dispersion, it is the first time that this problem  is proven to be polynomially solvable in 1D.
%  The properties of these DP algorithms are discussed for an efficient implementation, as well as the perspectives to extend these results for PFs in dimension three (3D) and more.

The remainder of this paper is organized as follows.
In Section 2, we introduce the notation and formally describe the problems.
In Section 3, we discuss related state-of-the-art elements to situate our contributions.
 In Section 4, intermediate results are presented. %, allowing to define the Max-Sum-Neighbor $p$-dispersion problem.
 In Sections 5, 6 and 7,   DP algorithms with a polynomial complexity are respectively presented for the Max-Sum-Neighbor,
Max-min and Max-min-Sum  variants.
In Section 8, our results are discussed from both a theoretical and a practical point of view.
In Section 9,  our contributions and the open perspectives are summarized. %,  discussing also future directions of research.
% {To ease the readability of the paper, the elementary proofs of some intermediate results are gathered in Appendix A when they do not contain
% algorithmic elements that will be discussed.
% A variant of  DP algorithm   for the Max-min-Sum is presented in Appendix B.
%}

\section{Problem statement}

Let  $E=\{x_1,\dots, x_n\}$ be a set of $n$ points in a 2D PF, considering the minimization of two objectives (this is not a loss of generality).
We denote discrete intervals $[\![a,b]\!] = [a,b] \cap \ZZ$, so that we can use the notation of discrete index sets and write $E=\{x_i\}_{i \in [\![1,n]\!]}$.
% A 2D PF can be defined as follows: there exists no pair of points in $E$ such that one has lower or equal values for all the coordinates (or criterion). This defines comparability  zones  in Figure \ref{defIllustr}.
 {As in \cite{dupin2019medoids,dupin2021unified},  PFs are formally defined  using binary relations:  relation $\mathcal{I}$ expresses Pareto incomparability,  whereas  relation $ \prec$  defines an order from  left to  right, as illustrated in Figure \ref{defIllustr}.}
Binary relations $\mathcal{I},\prec $ are  defined  for all $ y=(y^1,y^2),z=(z^1,z^2) \in \RR^2$ with:
\begin{eqnarray}
 y \prec z  & \Longleftrightarrow  & y^1< z^1 \phantom{2} \mbox{and}\phantom{2} y^2> z^2\\
y \preccurlyeq z  & \Longleftrightarrow  &  y \prec z  \phantom{2} \mbox{or}\phantom{2} y= z\\
y \phantom{1}\mathcal{I}\phantom{1} z  & \Longleftrightarrow  &y \prec z  \phantom{2} \mbox{or} \phantom{2}  z \prec y .
\end{eqnarray}

  \begin{figure}[ht]
      \centering
   \begin{tikzpicture}[scale=4.25]
    %\draw[very thin,color=gray] (-1,-1) grid (1,1);
    \draw[->] (-1.2,0) -- (1.2,0) node[right] {$x$};
    \draw[->] (0,-0.5) -- (0,0.55) node[above] {$y$};
    \draw (0.3,0) node[above] {$O=(x_O,y_O)$};
    \draw (0,0) node[color=black%blue!50
    ] {$\bullet$};
    \draw (0.6,0.35) node[above] {$B=(x_B,y_B)$ with};
    \draw (0.6,0.25) node[above] {$x_O<x_B$ and $y_O<y_B$ };
    \draw (0.6,0.15) node[above] {$O$ dominates $B$};

    %\draw (0.3,0) node[above] {$A=(x_O,y_O)$};
    \draw (-0.6,-0.25) node[above] {$C=(x_C,y_C)$ with};
    \draw (-0.6,-0.35) node[above] {$x_O>x_C$ and $y_O>y_C$ };
    \draw (-0.6,-0.45) node[above] {$C$ dominates $O$};

    \draw (0.6,-0.25) node[above] {$D=(x_D,y_D)$ with};
    \draw (0.6,- 0.35) node[above] {$x_O<x_D$ and $y_D<y_O$ };
    \draw (0.6,- 0.45) node[above] {$O  \mathcal{I}  D \phantom{1}$ and $ \phantom{1} O \prec D$};
     % \draw (0.6,- 0.5) node[above] {$O$ and $D$ without dominance relation};
    %\draw (0.3,0) node[above] {$A=(x_O,y_O)$};
    \draw (-0.6,0.35) node[above] {$A=(x_A,y_A)$ with};
    \draw (-0.6,0.25) node[above] {$x_O>x_A$ and $y_O<y_A$ };
    \draw (-0.6,0.15) node[above] {$O  \mathcal{I}  A\phantom{1}$ and $\phantom{1} A \prec O$};
%      \draw (-0.6,0.2) node[above] {$A$ and $O$ without dominance relation};

\end{tikzpicture}
    \caption{Illustration of relations $\mathcal{I},\prec$ and  Pareto dominance minimizing two objectives indexed by $x$ and $y$:}   \label{defIllustr}
   \end{figure}

A 2D PF can be the projected costs of efficient solutions using exact approaches in discrete MOO problems  \cite{ehrgott2003multiobjective},
or using population meta-heuristics like an evolutionary algorithm (EA) \cite{talbi2009metaheuristics}.
In the context of databases, Skyline operators are also PFs   \cite{borzsony2001skyline}
A 2D PF can be extracted from any subset of $\RR^2$ using an output-sensitive algorithm \cite{nielsen1996output}.
MOO problems with  continuous variables may have as solution a continuous PF. %Using the results of this paper in this last case
We will discuss in Section \ref{sec::continuousPF} how to use the results of this paper in this last case.
Finally, an affine 2D PF is similar to a 1D instance, we will formalize it.

\vskip 0.23cm
A strong assumption in this paper is that the 2D PF $E$ is known a priori: p-dispersion problems are computed knowing $E$.
This version of the problem is denoted \emph{explicit}, unlike the so-called \emph{implicit} versions where points of the PF are selected by  simultaneously calculating the PF.
In the context of MOO optimization, \emph{implicit} p-dispersion would be combined with MOO approaches as in \cite{bazgan2017discrete}.
Context of databases are less structured: an \emph{implicit} version of p-dispersion would consider the total number of rows in the database $h \gg n$ that can be enumerated where $n$ is the size of the Skyline, which is unkown at the beginning of the search, as in  \cite{cabello2020faster}.

\vskip 0.23cm

The distance between points $x_i,x_j \in E$ is denoted %, we consider
 $d_{ij}=d(x_i,x_j)^{\alpha}$ where
  $\alpha>0$ and $d(y,z)$ denotes  a Chebyshev or a Minkowski distance,
 induced by the  $\ell_{\infty}$ and $\ell_m$ norms.
For a given $m>0$, Minkowski distance $d_m$ is defined  by the following formula for $ y=(y^1,y^2),z=(z^1,z^2) \in \RR^2$:
\begin{equation}\label{distEucl}
\forall y,z \in \RR^2, \;\;\;
d_m(y,z) =  \sqrt[m\,]{ \left|y^1 - z^1\right|^m + \left|y^2 - z^2\right|^m}.
\end{equation}
The case $m=2$ corresponds to the Euclidean distance.
The limit with $m \rightarrow \infty$ defines the Chebyshev distance, denoted $d_{\infty}$:
%and given by the formula: 
\begin{equation}\label{distCheb}
\forall y,z %=(y^1,y^2),z=(z^1,z^2)
\in \RR^2, \;\;\;
d_{\infty}(y,z) =  \max\left( \left|y^1 - z^1\right|,\left|y^2 - z^2\right|\right).
\end{equation}

\vskip 0.23cm

In $p$-dispersion problems, the task is to
select $p\geqslant 2$ out of $n$ given candidate points, while maximizing 
a dispersion function $f$:
% measuring the dispersion of the points. 
% Generally discrete p dispersion problems are in the shape:
\begin{equation} \label{pDispersionDefGen}
\PP_{disp}(E,p)=  \max_{(z_1,z_2,\dots, z_p) \in D_p}  \phantom{2}
 f(z_1,z_2,\dots, z_p),
 \end{equation}
 where $D_p$ denotes the set of all the p-tuples with distinct points of $E$:
\begin{equation} 
D_p=  \{(z_1,z_2,\dots, z_p) \in E^p \phantom{1}|\phantom{1} \forall 1 \leqslant i < j \leqslant p,  z_i \neq z_j\}.
 \end{equation}

The most standard $p$-dispersion problem is also denoted Max-min $p$-dispersion problem or  $p$-dispersion-Mm.
The dispersion function  is in this case  the minimum
of distances  $d_{ij}$ between pairs of the selected points.
%Formally, t
 The Max-min $p$-dispersion problem for $p\geqslant 2$ is  written as:
 \begin{equation} \label{pDispersionDef}
\PP_{disp}^{Mm}(E,p)=  \max_{(z_1,z_2,\dots, z_p) \in D_p}\phantom{2}
 \min_{i,j : 1 \leqslant i < j \leqslant p }  d_{ij}.
 \end{equation}
%Another well known variant is the
Max-Sum(-Sum) dispersion problem, denoted $p$-dispersion-MS, considers  as dispersion function
the total sum of distances among selected points:
%with:
 \begin{equation} \label{pmaxSumDispersionDef}
\PP_{disp}^{MS}(E,p)=  \max_{(z_1,z_2,\dots, z_p) \in D_p}  
 \phantom{2} \sum_{i=1}^{p-1} \sum_{j=i+1}^p  d_{ij}.
 \end{equation}

 We consider also  the  Max-Sum-min  $p$-dispersion variant, denoted $p$-dispersion-MSm \cite{erkut1991comparison},
where dispersion is measured as %dispersion objective
the sum of the distance of each selected points to its closest (and different) selected point:
   \begin{equation} \label{pmaxSumMinDispersionDef}
%\forall \alpha >0, \;\;\;
\PP_{disp}^{MSm}(E,p)=  \max_{(z_1,z_2,\dots, z_p) \in D_p} \phantom{2}
 \sum_{i=1}^{ p }  \min_{j \in [\![1,p]\!]-\{i\}  } d_{ij}.
 \end{equation}
% Another specific variant of Max-Sum-min $p$-dispersion in a 2D PF will be defined in the Section 4, using specific properties of a 2D PF.
 The Max-min-Sum  $p$-dispersion problem \cite{erkut1991comparison},  denoted $p$-dispersion-MmS, is defined
with a dispersion calculated as %dispersion objective
the sum of distances of each selected point to the other selected points:
 \begin{equation} \label{pmaxMinSumDispersionDef}
%\forall \alpha >0, \;\;\;
\PP_{disp}^{MmS}(E,p)=  \max_{(z_1,z_2,\dots, z_p) \in D_p} \phantom{2}
 \min_{i\in [\![1,p]\!]}  \sum_{j \in [\![1,p]\!]-\{i\} } d_{ij}.
 \end{equation}

 \vskip 0.24cm
Note that 1D instances are special cases of 2D PF, equivalent to aligned points in a 2D PF. Considering any variant of $p$-dispersion problems, only the relative distance matters. Hence, $p$-dispersion problems for aligned points in the plane are equivalent to 1D instances and in these cases, Minkowski and Chebyshev distances are the same.

 The previous  definitions are generic and do not use specificities of 2D PFs. Lemmas  \ref{reord}  and \ref{orderDist}, mentioned  and proven in \cite{dupin2021unified}, allows to reformulate the definitions
of some $p$-dispersion variants in a 2D PF, as well as defining a new variant that will be specific for 2D PFs:
%with a re-indexing in  $O(n\log n)$ time.

\begin{lemma}%[Total order in a 2D PF]
\label{reord}
Let $E=\{x_i\}_{i \in [\![1,n]\!]}$ be a 2D PF.
The relation $\preccurlyeq$ is an order  and $\prec$ is  transitive.
$E$ can be re-indexed in  $O(n\log n)$ time  such that:
\begin{eqnarray}
\forall (i_1,i_2) \in [\![1,n]\!]^2, & i_1<i_2 \Longrightarrow & x_{i_1} \prec x_{i_2}\label{ordCroissant}, \\
\forall (i_1,i_2) \in [\![1,n]\!]^2, &  i_1\leqslant i_2 \Longrightarrow & x_{i_1} \preccurlyeq x_{i_2}. \label{ordCroissant2}
 \end{eqnarray}
\end{lemma}
\begin{lemma}%[Total order and monotony]
\label{transitiv}\label{orderDist}
Let $E=\{x_i\}_{i \in [\![1,n]\!]}$ be a re-indexed 2D PF usinf Lemma  \ref{reord}.
The following monotony relations are valid considering
 a Minkowski or Chebyshev distance $d$, and any real number $\alpha >0$:
%Let $(i_1,i_2,i_3) \in [\![1,n]\!]^3$.
\begin{eqnarray}
\forall (i_1,i_2,i_3) \in [\![1,n]\!]^3, \;\;\;&   i_1 \leqslant i_2<i_3 \Longrightarrow
d(x_{i_1},x_{i_2})^{\alpha}  < d(x_{i_1},x_{i_3})^{\alpha}  \label{eq1} \\
\forall (i_1,i_2,i_3) \in [\![1,n]\!]^3, \;\;\; &   i_1<i_2 \leqslant i_3 \Longrightarrow
d(x_{i_2},x_{i_3})^{\alpha}  < d(x_{i_1},x_{i_3})^{\alpha} \label{eq2}
\end{eqnarray}
\end{lemma}

\vskip 0.23cm
Lemma \ref{reord}   defines a 1D structure and a total order in a 2D PF.
The  re-indexing in  Lemma  \ref{reord} is equivalent to a lexicographic sort minimizing hierarchically the two objectives, thus running in$O(n\log n)$ time.
We will not use  Lemma  \ref{reord}  when the $O(n\log n)$  time complexity degrade the overall complexity of some algorithms.
When the indexing of a 2D PF fulfill equations (\ref{ordCroissant}),  it will be mentioned as a \emph{re-indexed 2D PF}.
Without additional precision, a 2D PF will not be considered as re-indexed with  Lemma  \ref{reord}.

\vskip 0.23cm

  \begin{figure}[ht]
      \centering
   \begin{tikzpicture}[scale=0.5]
    %\draw[very thin,color=gray] (-1,-1) grid (1,1);
    \draw[->] (0,0) -- (20,0) node[right] {$\emph{Obj}_1$};
    \draw[->] (0,0) -- (0,10.6) node[above] {$\emph{Obj}_2$};
    \draw (1.2,9.7) node[above] {$x_1$};
    \draw (1,9.7) node[color=blue!50] {$\bullet$};
    \draw (1.4,8.1) node[above] {$x_2$};
    \draw (1.7,8.1) node[color=blue!50] {$\bullet$};
    \draw (2.7,6.8) node[above] {$x_3$};
    \draw (2.5,6.8) node[color=blue!50] {$\bullet$};
    \draw (3.7,6.4) node[above] {$x_4$};
    \draw (3.5,6.4) node[color=blue!50] {$\bullet$};
    \draw (4.7,6.1) node[above] {$x_5$};
    \draw (4.5,6.1) node[color=blue!50] {$\bullet$};
    \draw (5.2,5.3) node[above] {$x_6$};
    \draw (5,5.3) node[color=blue!50] {$\bullet$};
    \draw (6.5,5.1) node[above] {$x_7$};
    \draw (6.3,5.1) node[color=blue!50] {$\bullet$};
    \draw (8.2,4.7) node[above] {$x_8$};
    \draw (8,4.7) node[color=blue!50] {$\bullet$};
    \draw (9.02,3.4) node[above] {$x_9$};
    \draw (9,3.4) node[color=blue!50] {$\bullet$};
    \draw (10.2,3.1) node[above] {$x_{10}$};
    \draw (10,3.1) node[color=blue!50] {$\bullet$};
    \draw (11.42,2.7) node[above] {$x_{11}$};
    \draw (11,2.7) node[color=blue!50] {$\bullet$};
    \draw (12.9,2.4) node[above] {$x_{12}$};
    \draw (12.7,2.4) node[color=blue!50] {$\bullet$};
    \draw (14.7,2.1) node[above] {$x_{13}$};
    \draw (14.3,2.1) node[color=blue!50] {$\bullet$};
    \draw (16.6,1.4) node[above] {$x_{14}$};
    \draw (16.4,1.4) node[color=blue!50] {$\bullet$};
    \draw (18.9,0.9) node[above] {$x_{15}$};
    \draw (18.7,0.9) node[color=blue!50] {$\bullet$};
    \end{tikzpicture}
    \caption{Illustration of a 2D PF with 15 points and the indexing implied by Lemma \ref{reord} }\label{orderIllustr}
   \end{figure}

Using Lemma \ref{orderDist},  Max-min and Max-Sum-min $p$-dispersion problems can be reformulated in a 2D PF considering only consecutive distances: % (the full proof is in Appendix A) :}
\begin{lemma}\label{reformulation}
Let $E=\{x_i\}_{i \in [\![1,n]\!]}$ be a re-indexed 2D PF.
%Let  $E =\{x_1,\dots, x_n\}$ a subset of $n$ points  of $\RR^2$, %fulfilling hypothesis (\ref{hypoNonDominated})
%such that for all $ i\neq j$, $x_i \prec x_j$.
The Max-min  and Max-Sum-min $p$-dispersion problems in $E$ are also defined as:
% \begin{eqnarray}
% \PP_{disp}^{Mm}(E,p)&=&  \max_{1 \leqslant i_1<\dots< i_p \leqslant p }   \min_{j \in [\![ 1 ;  p-1]\!] }  d_{i_j,i_{j+1}}\label{pDispersion2DPF}\\
% \PP_{disp}^{MSm}(E,p)&= & \max_{1 \leqslant i_1<\dots< i_p \leqslant p }   \sum_{j=2}^{ p-1 } \min( d_{i_j,i_{j+1}},d_{i_j,i_{j-1}}) + d_{i_1,i_{2}} + d_{i_{p-1},i_p}  \label{pmaxSumMinDispersionDefNew}
%  \end{eqnarray}
\begin{equation} \label{pDispersion2DPF}
\PP_{disp}^{Mm}(E,p)=  \max_{1 \leqslant i_1<\dots< i_p \leqslant p }  \phantom{1}
 \min_{j \in [\![ 1 ;  p-1]\!] }  d_{i_j,i_{j+1}}
 \end{equation}
\begin{equation} \label{pmaxSumMinDispersionDefNew}
\PP_{disp}^{MSm}(E,p)=  \max_{1 \leqslant i_1<\dots< i_p \leqslant p }  \phantom{1}
 \sum_{j=2}^{ p-1 } \min( d_{i_j,i_{j+1}},d_{i_j,i_{j-1}}) + d_{i_1,i_{2}} + d_{i_{p-1},i_p}
 \end{equation}
\end{lemma}

    \noindent{\textbf{Proof}}: In the inner minimization of
(\ref{pDispersionDef}) and (\ref{pmaxSumMinDispersionDef}), distances
$d_{i_j,i_{j'}}$ are considered. Such distances are higher than  $d_{i_j,i_{j+1}}$ and $d_{i_{j'-1},i_{j'}}$ using Lemma \ref{orderDist}, and it remains only distances among consecutive points.
For the two extreme points $x_1$ and $x_p$, it remains only
 $d_{i_1,i_{2}}$ and $d_{i_{p-1},i_p}$. % in the last minimization of
%(\ref{pDispersionDef}) and (\ref{pmaxSumMinDispersionDef}).
$ \hfill\square$

% \noindent{\textbf{Proof}}: In the inner minimization of
% (\ref{pDispersionDef}) and (\ref{pmaxSumMinDispersionDef}), distances
% $d_{i_j,i_{j'}}$ are considered. Such distances are higher than  $d_{i_j,i_{j+1}}$ and $d_{i_{j'},i_{j'+1}}$ using Lemma \ref{orderDist}, and it remains only distances among consecutive points. % in the last minimization of
% %(\ref{pDispersionDef}) and (\ref{pmaxSumMinDispersionDef}).
% $ \hfill\square$

\vskip 0.24cm
{Furthermore, Lemma \ref{reord} allows to define a new variant of Max-Sum-min $p$-dispersion, the Max-Sum-Neighbor $p$-dispersion problem:} % or  $p$-dispersion-MSN:
% ,
% and using the order of Proposition \ref{reord} to have a notion of neighbors and closest neighbors:

\begin{definition}[Max-Sum-Neighbor $p$-dispersion]
{Let $E=\{x_i\}_{i \in [\![1,n]\!]}$ be a re-indexed 2D PF.
 Max-Sum-Neighbor $p$-dispersion,  denoted $p$-dispersion-MSN, is defined in  $E$ summing
 only the distances between neighbor points:
    \begin{equation} \label{pmaxSumMinDispersionDef2}
\PP_{disp}^{MSN}(E,p)=  \max_{1 \leqslant i_1<i_2<\dots< i_p \leqslant p}
 \sum_{j=1}^{ p-1 }   d_{i_j,i_{j+1}}
 \end{equation}
 }
\end{definition}
%\vskip 0.2cm

\begin{figure}[ht]
\caption{{Illustration of the different dispersion variants: we consider the 4-dispersion problems, with four selected points, A,B,C and D such that $A \prec B \prec C \prec D$.}}\label{illustr4Dispersion}

      \includegraphics[angle=0, width=.85\linewidth]{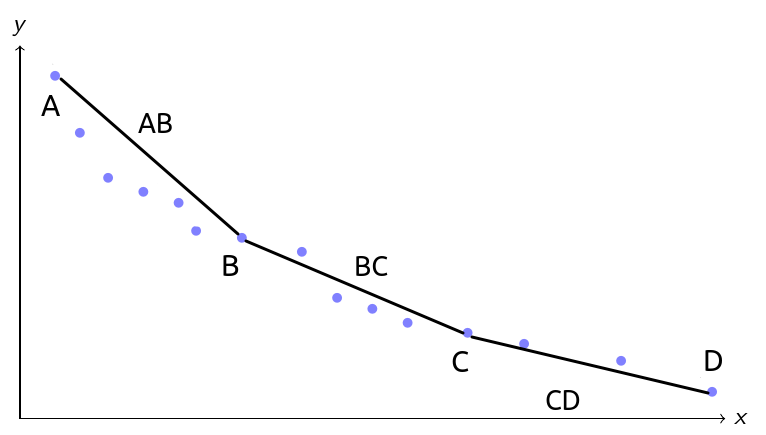}

      \vskip 0.42cm

\begin{tabular}{ l c l}
\hline
Variant && Dispersion of A,B,C,D\\
\hline
%$p$-dispersion-
Mm&:& $\min$(AB$^{\alpha}$,BC$^{\alpha}$,CD$^{\alpha}$)$ = \min$(AB,BC,CD)$^{\alpha}$\\
%$p$-dispersion-
MS&:& AB$^{\alpha}$+AC$^{\alpha}$+AD$^{\alpha}$+BC$^{\alpha}$+BD$^{\alpha}$+CD$^{\alpha}$\\
%$p$-dispersion-
MSN&:& AB$^{\alpha}$+BC$^{\alpha}$+CD$^{\alpha}$\\
%$p$-dispersion-
MSm&:& AB$^{\alpha}$+min(AB$^{\alpha}$,BC$^{\alpha}$)+min(BC$^{\alpha}$,CD$^{\alpha}$)+CD$^{\alpha}$\\
%$p$-dispersion-
MmS&:& min(AB$^{\alpha}$+AC$^{\alpha}$+AD$^{\alpha}$, AB$^{\alpha}$+BC$^{\alpha}$+BD$^{\alpha}$,\\
&&AC$^{\alpha}$+BC$^{\alpha}$+CD$^{\alpha}$, AD$^{\alpha}$+BD$^{\alpha}$+CD$^{\alpha}$)\\
\hline
\end{tabular}
\end{figure}

Dispersion variants are illustrated in Figure \ref{illustr4Dispersion}.
 Lemma \ref{reformulation} shows that $p$-dispersion-MSm is not a symmetric expression of distances, inducing more importance to extreme distances AB and CD.
On the contrary, $p$-dispersion-MSN induces symmetrical expressions.%, as illustrated in in Figure \ref{illustr4Dispersion}. %This motivates the introduction of $p$-dispersion-MSN.

% $k$-medoids?

\section{Related works}
%
% This section describes   related results and approaches in order to appreciate the contributions of this paper.

\subsection{Complexity results for $p$-dispersion problems}

Max-min and Max-Sum $p$-dispersion problems are $\mathcal{N}\mathcal{P}$-hard for general metric spaces.
This is proven in both cases by a polynomial reduction from the maximum independent set problem \cite{erkut1990discrete,hansen1995dispersing}.
Max-min and Max-Sum $p$-dispersion problems are still $\mathcal{N}\mathcal{P}$-hard problems when distances fulfill the triangle inequality \cite{erkut1990discrete,hansen1995dispersing}.
The planar  Max-min $p$-dispersion problem is also $\mathcal{N}\mathcal{P}$-hard \cite{wang1988study}, the $\mathcal{N}\mathcal{P}$-hardness of the planar Max-Sum $p$-dispersion problem is still an open question to our knowledge.

Approximability and non approximability with constant factors were studied for  $p$-dispersion problems.
Unless $\mathcal{P}=\mathcal{N}\mathcal{P}$,   Max-min $p$-dispersion  cannot be approximated with a constant factor \cite{ravi1994heuristic}.
For general metric spaces,   Max-min and   Max-Sum $p$-dispersion problems can be approximated with a 1/2 factor \cite{tamir1991obnoxious,tamir1998comments}.
For 2D spaces,   Max-min and   Max-Sum $p$-dispersion problems can be approximated with a constant factor \cite{ravi1994heuristic}.
The 1/2 factor is the tightest approximation factor for Max-min $p$-dispersion with triangle inequality,  unless $\mathcal{P}=\mathcal{N}\mathcal{P}$  \cite{ravi1994heuristic}.
%Other lower bounds on the approximation factor are still open questions \cite{ravi1994heuristic}.
  
Some sub-cases of  $p$-dispersion problems are solvable in polynomial time.
The 1D cases of Max-min and Max-Sum $p$-dispersion problems
are solvable in polynomial time, with  DP algorithms running in $O(\max\{pn,n \log n\})$  time \cite{ravi1994heuristic,wang1988study}.
%A duality theorem holds between Max-min $p$-dispersion and $p$-center problems on the line and in tree structures: Max-min $p$-dispersion   is the dual of the (continuous) (p-1)-center \cite{shier1977min}.
%Max-min $p$-dispersion is solvable in 1D  in $O(n \log n)$ time \cite{chandrasekaran1982polynomially,megiddo1983new}.
Within a tree structure,  Max-min $p$-dispersion is solvable  in $O(n^2 \log n)$ time \cite{chandrasekaran1981location}.
 Although,  $p$-dispersion is mentioned to have relevant applications for MOO \cite{erkut1991comparison,ravi1994heuristic}, no specific studies concerned  $p$-dispersion in a PF besides 1D cases to the best of our knowledge.
%Several results concern 1D instances, which is equivalent to the sub-case of aligned points in a 2D PF.
%For 1D instances, Max-min and Max-Sum $p$-dispersion problems are solvable in $O(\max\{pn,n \log n\})$  time  \cite{ravi1994heuristic,wang1988study}.
 We note an interest for implicit versions of  dispersion problems in of Skyline Operators \cite{lin2007selecting,magnani2014taking,valkanas2013skydiver}.

\subsection{Exact methods for $p$-dispersion problems}

 Max-Sum $p$-dispersion  can be formulated as a quadratic optimization problem, defining   binary variables $z_{j} \in \{0,1\}$   with
$z_{j}=1$ if and only if the point $x_j$ is selected:
\begin{equation} \label{mathProgMaxSumDispersion}
  \begin{array}{lrrlr}
 \PP_{disp}^{MS}(E,p)&=  \max &\displaystyle \sum_{i=1}^n\sum_{j=i+1}^n d_{i,j} z_i z_j  && (\ref{mathProgMaxSumDispersion}.1)\\% \displaystyle \sum_{1\leqslant i<j\leqslant N} d_{ij}z_i z_j \\
 & \mbox{s.t.:} & \sum_{j=1}^n z_{j}  = p && (\ref{mathProgMaxSumDispersion}.2)\\
  & & z_{j} \in \{0,1\} & \hskip 0.2cm   \forall j \in [\![1,n]\!].& (\ref{mathProgMaxSumDispersion}.3)
  \end{array}
\end{equation}
Linearizing  (\ref{mathProgMaxSumDispersion}.1) leads to the 
 Integer Linear Programming (ILP) formulation %for the $p$-dispersion problem
 provided in \cite{kuby1987programming}.
Exact  Branch\&Bound (B\&B) algorithms were also provided  computing iteratively higher and lower bounds, with a
 Lagrangian relaxation  \cite{augca2000lagrangian}
or with tailored higher bounds computable in $O(n^3)$ time   \cite{pisinger2006upper}.

Max-min $p$-dispersion  is also a non linear optimization problem \cite{pisinger2006upper}:
\begin{equation} \label{mathProgDispersion}
  \begin{array}{lrlrr}
 \PP_{disp}^{Mm}(E,p)= &\displaystyle \max_{d \geqslant 0} & d && (\ref{mathProgDispersion}.1)\\% \displaystyle \sum_{1\leqslant i<j\leqslant N} d_{ij}z_i z_j \\
 &\mbox{s.t:}  & \sum_{j=1}^n z_{j}  = p && (\ref{mathProgDispersion}.2)\\
%    & \sum_{j=1}^n x_{i,j} &= 1 &   \forall i \in [\![1,n]\!] \\
 && d z_i z_j \leqslant  d_{i,j}&  \hskip 0.2cm  \forall 1\leqslant i<j\leqslant n, & (\ref{mathProgDispersion}.3) \\
  & & z_{j} \in \{0,1\}&  \hskip 0.2cm   \forall j \in [\![1,n]\!].& (\ref{mathProgDispersion}.4)
  \end{array}
\end{equation}
% where the  binary variables $z_{j} \in \{0,1\}$  are defined with 
% $z_{j}=1$ if and only if the point $x_j$ is selected.
 The standard linearization of constraints (\ref{mathProgDispersion}.3) leads to the 
Mixed Integer Linear Programming (MILP) formulation %for the Max-min $p$-dispersion problem also
 \cite{kuby1987programming}.
%Sayah and Irnich proposed 
 An alternative  MILP formulation  and specific cuts for a Branch\&Cut algorithm
 is provided by \cite{sayah2017new}.
Decomposition schemes from \cite{augca2000lagrangian} and \cite{pisinger2006upper}
have  been also extended for the Max-min $p$-dispersion problem.
% Recently,  Max-min $p$-dispersion problem was solved by
% 
% a decremental clustering method 
% to reduce the problem to the solution of a series of smaller
% pDPs until reaching proven optimality. A $k$-means algorithm
% 
%   
%   TODO:  memory/time limits may prevent thesolution of problems containing more than a few hundred nodes
% 
%   \cite{contardo2020decremental}
%   title={Decremental clustering for the solution of $p$-dispersion problems to proven optimality},

  Similarly, MILP formulations were designed for 
the  Max-Sum-min and Max-min-Sum $p$-dispersion variants \cite{erkut1991comparison}.
Such variants were less studied. A recent work proposed a unified MILP formulation and B\&B algorithm for the
four variants of $p$-dispersion problems \cite{lei2015unified}.

% \subsection{Heuristics to solve $p$-dispersion problems}
% 
% \cite{erkut1994comparison}
% 
% 
% Glover et al. (1998), Silva et al. (2004),
% and Duarte and Martıi(2007).
% 
% max sum: Tabu search GRASP
% 
% \cite{glover1998heuristic,resende2010grasp,marti2013heuristics}
% 
% 
% 
% Because of the flat landscape of Max-min problems, these papers agree that the
% MmDP presents a challenge to solution methods based on heuristic optimization.
% 
% Meta-heuristics \cite{resende2010grasp}
%  
%   maximum independent set
%  

\subsection{Clustering/selecting points in PFs}

We summarize here results related to the explicit versions of selection or  clustering of points in a PF.

Maximizing the quality of discrete representations of Pareto sets was studied with   hypervolume measure
in the Hypervolume Subset Selection (HSS) problem \cite{auger2009investigating,sayin2000measuring}. 
 The HSS problem, maximizing  representativeness of $k$ solutions among a PF of size $n$,  is $\mathcal{N}\mathcal{P}$-hard in 3D (and higher dimensions)  \cite{bringmann2018maximum}.
An exact  algorithm in  $n^{O(\sqrt{k})}$ {in 3D} and a  polynomial-time approximation scheme for any constant dimension $d$ were also provided  \cite{bringmann2018maximum}.
2D PF instances are solvable in polynomial time,  a  DP algorithm running
 in $O(kn^2)$ time and using $O(kn)$ space was firstly provided in \cite{auger2009investigating}.
The time complexity  was improved in $O(kn+n \log n)$ by \cite{bringmann2014two} and in $O(k(n-k)+n \log n)$ by \cite{kuhn2016hypervolume}.
Some similar results exist also for clustering problems.
The $k$-median and $k$-medoid problems are $\mathcal{N}\mathcal{P}$ hard in  2D since \cite{megiddo1984complexity}.
The 2D PF cases are solvable in $O(n^3)$  time with DP algorithms \cite{dupin2018clustering,dupin2019medoids}.
The 1D cases  are solvable in  $O(nk)$ time \cite{hassin1991improved}.
Min Sum of Square Clustering (MSSC), also denoted $k$-means problem,
is also $\mathcal{N}\mathcal{P}$-hard for 2D cases \cite{mahajan2012planar}.
% The restriction to 2D PF is conjectured in \cite{dupin2018dynamic} to  be  solvable in $O(n^3)$ time with a DP algorithm.
The 1D cases of $k$-means are also solvable by a DP algorithm,
with a complexity in $O(kn)$ using memory  space in $O(n)$ \cite{gronlund2017fast}.
%  For $k$-means, $k$-medoids, $k$-median,  significant differences in the  complexities exist  between the 2D PF and 1D cases.

 Similar results are  available for variants of $p$-center problems.
 The $p$-center problems minimize the radius of a ball, to cover all the points  with  $p$ such identical balls.
Contrary to the continuous version,  discrete $p$-center variant consider the additional constraint to have  a discrete set of points as candidates for ball centers. It makes sense to consider the original points as such candidates for centers, as in \cite{dupin2021unified}.
 The dual of a Max-min $p$-dispersion is similar to a min-max optimization as in $p$-center problems.
Duality relations hold between  $p$-dispersion-Mm and $p$-center problems
 in 1D and in tree structures:   $p$-dispersion-Mm   is the dual of the continuous $(p-1)$-center for such cases \cite{shier1977min}.
 Max-min $p$-dispersion and $p$-center problems have similar complexity results.
 The discrete and continuous $p$-center problems are $\mathcal{N}\mathcal{P}$-hard in general,
the  discrete $p$-center problem in $\RR^2$ with a Euclidean distance is also $\mathcal{N}\mathcal{P}$-hard \cite{megiddo1984complexity}.
The 1D and 2D PF sub-cases of $p$-center problems are polynomially solvable with DP algorithms.
The 1D  continuous $p$-center is solvable in $O(n \log^3 n)$ time  \cite{megiddo1983new},
whereas the time complexity in a 2D PF is in  $O(pn \log n)$  \cite{dupin2019planar,dupin2021unified}.
The discrete $p$-center problem in a 2D PF is solvable  in $O(n \log n)$ time and $O(n)$ space \cite{cabello2020faster},
whereas it is solvable in $O(n)$ time in 1D  \cite{frederickson1991parametric}.
The Min-sum $p$-radii problems, denoted also min-sum-diameter, are $p$-center variants, where the covering balls may  not be  identical. It is a min-Sum-Max optimization, that we can compare to the
Max-Sum-min optimization with $p$-dispersion-MSm.
 Min-sum $p$-radii  is $\mathcal{N}\mathcal{P}$-hard in the general case and polynomial within a tree structure \cite{doddi2000approximation}.
The $\mathcal{N}\mathcal{P}$-hardness is also proven even in metrics induced by weighted planar graphs
\cite{gibson2010metric}.
In a 2D PF, Min-sum $p$-radii  is solvable in  $O(pn^2)$ time with a DP algorithm  \cite{dupin2021unified}.
In 1D, a specific algorithm runs in $O(n \log n)$ time  \cite{dupin2021unified}.

Lastly, partial variants were studied for  $p$-center problems and variants in a 2D PF,
allowing that $m$ points are not covered. These points can be outliers to remove or isolated points that are wished to be detected in the context of EAs \cite{dupin2021unified}.
% Detecting  isolated points is useful for operators  of  population MOO meta-heuristics to focus on  zones around the isolated points to have a more balanced PF representation.
DP algorithms for 2D PF can be extended allowing to uncover $m>0$ points. The time and space complexity of DP algorithms are
then multiplied by a factor $m$ \cite{dupin2021unified}.
In particular, partial $p$-center problems in a 2D PF are solvable in  $O(mpn \log n)$ time and $O(mn)$ space \cite{dupin2021unified}.
%which is reasonable for an application inside   population MOO meta-heuristics.

% For $p$-center problems, the complexity for 2D PF problems is similar to the one from 1D cases, i.e. affine 2D PF.

\subsection{Summary of contributions and relations to state of the art} \label{sec::synthese}

From this literature review, some common results appear. General 2D instances of the clustering and selection problems are often $\mathcal{N}\mathcal{P}$-hard. DP algorithms induce a polynomial complexity in 1D and some 2D PF cases.
Many  DP algorithms use a $p \times n$ matrix to store results of  $k \leqslant p$ selection/clustering among the $n' \leqslant n$ first points. This induces $p \times n$  computations of a partial solution using $O(n)$ previous ones. A linear enumeration induces time complexities in $O(p  n^2)$
whereas some $O(p  n \log n)$ time complexities can be obtained using a logarithmic search.
Our DP algorithms for $p$-dispersion-MSN and $p$-dispersion-Mm have this form.
Other techniques make it possible to divide the complexity by a factor of  $p$, as in \cite{cabello2020faster,gronlund2017fast}.
It is an open perspective for the DP algorithms provided in this paper.

Table \ref{complexityTable} summarizes the complexity results for  selection and clustering problems for the 1D, 2D PF and 2D sub-cases. A first comparison between these problems situates the selection problems $p$-center, $p$-dispersion-Mm and HSS as the fastest to solve in 2D PFs.
 Clustering with $p$-median or $p$-medoids is more accurate for measuring cluster similarity, however the time complexity in  $O(n^3)$ is  a burden for application. Time complexity in $O(pn^3)$ for $p$-dispersion-MSm %is a bottleneck for application, which
  is a more a theoretical than a practical result.
  Surprisingly, $p$-dispersion-MSN, which seems to be a slight variant of $p$-dispersion-MSm, has a much better complexity.
Min-sum $p$-radii and $p$-dispersion-MSN  have the same time complexity in $O(p n^2)$ with similar DP algorithms.
Both variants  improve weaknesses of max-min and min-max optimization with $p$-center and $p$-dispersion-Mm, such as the possible large number of optimal solutions, including solutions potentially very unbalanced, to obtain fewer and better balanced solutions.
Such weaknesses of $p$-dispersion-Mm  are highlighted  in Proposition \ref{backtrackOrderLemma},  similar algorithms and results
 for p-center problems are described in  \cite{dupin2021unified}.
%Note that the complexity of Min-Sum $p$-dispersion for 2D and 2D PF is still an open question. %, whereas the 1D instances are polynomially solvable  \cite{wang1988study}.

  \begin{table}
\caption{{Comparison of the time complexity obtained after our study for dispersion problems (in bold) with related problems  on 1D, 2D PF and 2D cases and their reference. Some cases are still open questions. BF denotes brute force naive enumeration. Note that
$p$-dispersion-MSN is defined only for 1D and 2D PF instances, not for general 2D instances} }\label{complexityTable}
\begin{tabular}{|l|cc|cc|cc|}
\hline
problem & 1D  &&  2D PF  && 2D  & \\
\hline
$p$-dispersion-Mm & $O(n(p+ \log n))$  & \cite{ravi1994heuristic}  & $\mathbf{O(pn \log n)}$ &   & $\mathcal{N}\mathcal{P}$-hard & \cite{wang1988study}\\
$p$-dispersion-MSN& $\mathbf{O(pn^2)}$ &   & $\mathbf{O(pn^2)}$ & & not defined &  \\
$p$-dispersion-MSm& $\mathbf{O(pn^3)}$ &   & $\mathbf{O(pn^3)}$ & &open & \\
$p$-dispersion-MS& $O(n(p+ \log n))$ &  \cite{wang1988study}  &open & & open& \\
\hline
$2$-dispersion& $\mathbf{O(n)}$ &   & $\mathbf{O(n)}$ & &$O(n^2)$ &BF  \\
$3$-dispersion& $\mathbf{O(n)}$ &   & $\mathbf{O(n)}$ & &$O(n^3)$ &BF  \\
$4$-dispersion& $\mathbf{O(n^2)}$ &   & $\mathbf{O(n^2)}$ & & $O(n^4)$ &BF  \\
\hline
Cont. $p$-center & $O(n \log^3 n)$ & \cite{megiddo1983new}  & $O(pn \log n)$ & \cite{dupin2021unified}  & $\mathcal{N}\mathcal{P}$-hard & \cite{megiddo1984complexity}\\
Discr.  $p$-center &$O(n)$ &  \cite{frederickson1991parametric}  & $O(n \log n)$ & \cite{cabello2020faster}& $\mathcal{N}\mathcal{P}$-hard & \cite{megiddo1984complexity}\\
Min-sum $p$-radii& $O(n \log n)$ & \cite{dupin2021unified}  & $O(pn^2)$ & \cite{dupin2021unified}& $\mathcal{N}\mathcal{P}$-hard & \cite{gibson2010metric}\\
\hline
% Cont. 1-center & $O(N )$ & \cite{gonzalez1985clustering}  & $O(N)$ &\cite{dupin2021unified}  & $O(N)$ & \cite{gonzalez1985clustering}\\
% Discr. 1-center & - & - & $O(N)$ &\cite{dupin2021unified} & $O(n \log n)$ & \cite{brass2009computing}\\
 Cont. $2$-center & $O(n \log n)$ &\cite{dupin2021unified}  & $O(n \log n)$ &\cite{dupin2021unified}  & $O(n \log^2 n)$ & \cite{megiddo1983linear}\\
 Discr. $2$-center & $O(n \log n)$ &\cite{dupin2021unified} &$O(n \log n)$ & \cite{dupin2021unified} &  $O(n^{4/3}\log^5n)$ & \cite{agarwal1998discrete}\\
%partial 1-center  & - & - & $O(N\min(M,\log N))$ & Prop. \ref{partial1center} &$O(N^2 \log N)$ & \cite{drezner1981modified}\\
%Rect. 1-center& $O(N)$ & \cite{drezner1987rectangular}  & $O(N)$ & Prop.  \ref{propOptcPcent}& $O(N)$ & \cite{drezner1987rectangular} \\
%Rect. 2-center& $O(N)$ & \cite{drezner1987rectangular}  & $O(n \log n)$ & Prop.  \ref{2centers}& $O(N)$ & \cite{drezner1987rectangular} \\
\hline
HSS& undefined &   &  $O(n(p+ \log n))$ & \cite{kuhn2016hypervolume}& $\mathcal{N}\mathcal{P}$-hard & \cite{bringmann2018maximum} \\
$p$-means& $O(pn)$ & \cite{gronlund2017fast}  & open &  & $\mathcal{N}\mathcal{P}$-hard & \cite{mahajan2012planar} \\
$p$-median& $O(pn)$ & \cite{hassin1991improved}  & $O(n^3)$ &  \cite{dupin2018clustering}. & $\mathcal{N}\mathcal{P}$-hard & \cite{megiddo1984complexity} \\
$p$-medoids& $O(pn)$ & \cite{hassin1991improved}  & $O(n^3)$ &  \cite{dupin2019medoids}. & $\mathcal{N}\mathcal{P}$-hard & \cite{megiddo1984complexity} \\
\hline
\end{tabular}
%\vskip 0.2cm
%\begin{left}
% \small
%  - : problem not defined in this dimension\\
%  open : open question (these cases were not studied)\\
%  BF : brute force naive algorithm
% %\end{left}
% \begin{minipage}%[angle=0,width=.99\linewidth]
% %  - : problem not defined in this dimension
% %
% % open: open question (these cases were not studied)
% %
% % BF: brute force naive algorithm
% \end{minipage}
\end{table}

%One can also compare the complexity in the columns of Table \ref{complexityTable}.

Complexity results can be  significantly worse for 2D PFs than the 1D sub-cases, as for the Min-sum $p$-radii and $p$-medoids.
 In such cases, triangle inequality instead of additivity of distances  is crucial.
For $p$-center and $p$-dispersion problems, the time complexity for 2D PFs is not significantly worse than for 1D sub-cases.
Within a tree structure, which is another extension of 1D cases,   $p$-dispersion-Mm is solvable in $O(n^2 \log n)$ time \cite{chandrasekaran1981location} instead of $O(pn \log n)$ in 2D PF, the time complexity in a 2D PF is significantly better.
Note that Max-Sum-min $p$-dispersion was not studied before in 1D to the best of our knowledge, so that Theorem 3 and Corollary 1 proves for the first time that Max-Sum-min $p$-dispersion in 1D is solvable in polynomial time.
 Perspectives may be to improve this complexity result using distance additivity in 1D.

%   MSN in 1D is trivial with $\alpha = 1$

\section{Intermediate results} \label{sec::singleCluster}
   
This section presents intermediate results that will be a basis for future developments.
% Lemma \ref{orderDist} induces first  complexity results   in
%Propositions \ref{2Disp}, \ref{trivialCases3} and \ref{trivialCase4}.
 A key element is that the  extreme points of a re-indexed 2D PF are natural candidates for $p$-dispersion problems: %, as analyzed in Propositions \ref{2Disp} and \ref{extremePoints}:

% The following propositions try to generalize such results to higher dimensions. A first question is to determine if the extreme points ( $z_1$ and $z_n$ after re-indexing of Proposition \ref{reord} )
% are necessarily selected in an optimal solution, or at least if optimal solution exists with the selection of extreme points: 
% 
\begin{prop}[2-dispersion problems]
\label{2Disp}
{  Let $E=\{x_i\}_{i \in [\![1,n]\!]}$ be a 2D PF}.\\
2-dispersion problems are solvable in $O(n)$ time using $O(1)$ additional memory space in the 2D PF $E$, considering any variant of $p$-dispersion. %s (\ref{pDispersionDef}), (\ref{pmaxSumDispersionDef}), (\ref{pmaxSumMinDispersionDef}), (\ref{pmaxMinSumDispersionDef})  or (\ref{pmaxSumMinDispersionDef2}).
There is a unique optimal solution, selecting the extreme points $x_1$ and $x_n$.
\end{prop}

\noindent{\textbf{Proof}}:
Any 2-dispersion variant % (\ref{pDispersionDef}), (\ref{pmaxSumDispersionDef}), (\ref{pmaxSumMinDispersionDef}), (\ref{pmaxMinSumDispersionDef}) or (\ref{pmaxSumMinDispersionDef2}),
consider the same  problem: % is considered:
    \begin{equation} \label{2Disps}
\PP_2(E)=  \max_{1 \leqslant i<j \leqslant p}  d(x_i,x_{j}).
 \end{equation}
 Indeed,  $\PP_{disp}^{Mm}(E,2) = \PP_{disp}^{MS}(E,2)= \PP_{disp}^{MmS}(E,2) = \PP_{disp}^{MSN}(E,2) = \PP_2(E)$ and
 $\PP_{disp}^{MSm}(E,2) = 2 \PP_2(E)$.
%In the cases, (\ref{pmaxSumDispersionDef}), (\ref{pmaxSumMinDispersionDef}), the problem is actually $2 \PP_{disp}(E,2)$.
 Using Lemma \ref{orderDist}, $\PP_2(E)=d_{1,n}$,  selecting the two extreme points $x_1$ and $x_n$ after re-indexing.% , the optimal cost is $d_{1,n}$.
The complexity, once having computed $x_1$ and $x_n$ is in $O(1)$ time and additional space.
Re-indexing $E$ induces a complexity in $O(n \log n)$ time.
By computing only extreme points with a single traversal of $E$, %storing the current extreme points,
time complexity is in $O(n)$. $\hfill\square$

\vskip 0.23cm

\begin{prop}[$p$-dispersion and extreme points]
\label{extremePoints}
{  Let $E=\{x_i\}_{i \in [\![1,n]\!]}$ be a re-indexed 2D PF.}
For each $p$-dispersion variant, % (\ref{pDispersionDef}), (\ref{pmaxSumDispersionDef}), (\ref{pmaxSumMinDispersionDef}), (\ref{pmaxMinSumDispersionDef}) and (\ref{pmaxSumMinDispersionDef2}),
 an optimal solution %of $p$-dispersion
exists selecting $x_1$ and $x_n$ for $p\geqslant 2$. % once the points $(x_i)$ are re-indexed as in Proposition \ref{reord}.
Reciprocally, any optimal solution contains the extreme points in the case of the Max-Sum and Max-Sum-Neighbor variants.
\end{prop}

\noindent{\textbf{Proof}}:
%Considering variants (\ref{pDispersionDef}), (\ref{pmaxSumDispersionDef}), (\ref{pmaxSumMinDispersionDef}), (\ref{pmaxMinSumDispersionDef}) or (\ref{pmaxSumMinDispersionDef2}),
Let $1 \leqslant i_1<i_2<\dots< i_p \leqslant p$ be the  indexes defining an optimal solution of a $p$-dispersion variant.
Considering new indexes $i_1'=1, i_2' = i_2,\dots i_{p-1}'=i_{p-1}, i_p' = p$, Lemma \ref{orderDist} implies that  $d(x_{i_j},x_{i_{j'}})^{\alpha} \leqslant d(x_{i_j'},x_{i_{j'}'})^{\alpha}$ for all $j,j' \in [\![1,p]\!]$. Points  $x_{i_1'}, \dots ,x_{i_{p-1}'}, x_{i_{p}'}$ have at least the same dispersion than the original points $x_{i_1}, \dots ,x_{i_{p-1}}, x_{i_{p}}$. Hence, it defines an optimal solution of the considered   $p$-dispersion variant.
Having $1 < i_1$ or $i_p < p$, Lemma \ref{orderDist} induces %that we have furthermore
$d(x_{i_{1}},x_{i_{2}})^{\alpha} + d(x_{i_{p-1}},x_{i_{p}})^{\alpha} <
d(x_{i_{1}'},x_{i_{2}'})^{\alpha} + d(x_{i_{p-1}'},x_{i_{p}'})^{\alpha}$.
%d(i_{1}',i_{j_2}')^{\alpha}+ d(i_{p-1}',i_{p}')^{\alpha}$,
 Points  $x_{i_1'}, \dots ,x_{i_{p-1}'}, x_{i_{p}'}$ would induce a strictly better solution than an optimal solution of Max-Sum or Max-Sum-Neighbor $p$-dispersion.
By contradiction,  the extreme points are in any optimal solution for both variants.
$\hfill\square$

\vskip 0.23cm

 \noindent{\textbf{Remark}}:   For  Max-min $p$-dispersion,  optimal solutions exist without  containing  $x_1$ and $x_n$.
We  consider  3-dispersion-Mm  and 4 points $x_1 = (0,10)$,
$x_2 = (1,9)$,
$x_3 = (3,7)$,
$x_4 = (5,5)$. Using the Euclidean distance, $d(x_1,x_2)=\sqrt{2}$, $d(x_2,x_3)=2 \sqrt{2}$, $d(x_1,x_3)=3\sqrt{2}$, $d(x_3,x_4)=2 \sqrt{2}$.
Hence, $\{x_2,x_3,x_4\}$ has the same dispersion-Mm as $\{x_1,x_3,x_4\}$, $2 \sqrt{2}$ , which is optimal.
% is given with $x_1,x_2,x_3,x_4$ where $x_2,x_3,x_4$ are optimal regarding 3-dispersion without containing the extreme point $x_1$, considering
% Indeed, $d(x_1,x_2)=\sqrt{2}$, $d(x_2,x_3)=2 \sqrt{2}$, $d(x_1,x_3)=3\sqrt{2}$, $d(x_3,x_4)=2 \sqrt{2}$,
% and $x_2,x_3,x_4$ have the same dispersion as $x_1,x_3,x_4$, which is optimal.

\vskip 0.23cm
\noindent{\textbf{Remark}}:
Clustering measure like $k$-means, $k$-medoids or $k$-center variants do not return the extreme points in general.
% This property of $p$-dispersion problems is relevant in the context
% of  presenting a PF to a  decision maker.
% \cite{dupin2019medoids,dupin2021unified}.
\vskip 0.23cm
% \noindent{\textbf{Remark}}: Reciprocally

%Proposition \ref{extremePoints} allows to determine the complexity of $3$-dispersion problems:
%, and also improve the  general complexity enumerating all the possibilities, that would be in $O(n^p)$.

\begin{prop}[3-dispersion]\label{trivialCases3}
$3$-dispersion problems in a 2D PF are solvable in  $O(n)$ time using $O(1)$ additional space.
\end{prop}

\noindent{\textbf{Proof}}: Considering variants (\ref{pDispersionDef}), (\ref{pmaxSumDispersionDef}), (\ref{pmaxSumMinDispersionDef}), (\ref{pmaxMinSumDispersionDef}) or (\ref{pmaxSumMinDispersionDef2}),
we consider the two extreme points, which can be found in  $O(n)$ time
with one traversal of $E$. % like in Proposition \ref{2Disp}.
Then, there are $n-2$ cases to enumerate the last point, each cost computation of 3-dispersion being in $O(1)$ time, this last naive enumeration is in $O(n)$ time.
$3$-dispersion problems are thus solved in $O(n)$ time using $O(1)$ additional space with two traversals of $E$.$\hfill\square$
\vskip 0.23cm
\begin{prop}%[$p$-dispersion]
\label{trivialCase4}
 In a 2D PF, the  $p$-dispersion problems are solvable in  $O({p^2}$ ${n-2}\choose{p-2}$ $)$ time using $O(1)$ additional space.
\end{prop}

\noindent{\textbf{Proof}}: Similarly to Proposition \ref{trivialCases3}, once extreme points are computed in $O(n)$ time, the naive enumeration
of  other $p-2$ selected points induces ${n-2}\choose{p-2}$ computations, requiring $O(p)$ or $O(p^2)$ time computations.$\hfill\square$ 
 \vskip 0.23cm
 \noindent{\textbf{Remark}}: with $p \ll n$, the time complexity is roughly in $O(n^{p-2})$,
 instead of $O(n^{p})$ for the naive enumeration. Using Proposition \ref{trivialCase4},
cases  $p=4,5,6$ have respectively a time  complexity in $O(n^2)$, $O(n^3)$ and $O(n^4)$.
%In the following, complexities of these small $p$-dispersion problems may be improved only for specific cases using DP algorithms.
 \vskip 0.23cm
  A specific result holds for  $p$-dispersion-MSN in 1D.
%This problem was not studied and even defined before.
Proposition \ref{1DdispMSN} and its proof show that it makes sense to consider this problem only for  $\alpha \neq 1$:

\begin{prop}%[$p$-dispersion-MSN in  1D ]
\label{1DdispMSN}
  Let $E=\{x_i\}_{i \in [\![1,n]\!]}$ be a set of $n$ distinct real numbers, let $p \geqslant 2$.
If $\alpha =1$,
%Max-Sum-Neighbor
$p$-dispersion-MSN problem is solvable in $O(n)$ time. % using $O(1)$ additional memory space.
\end{prop}

\noindent{\textbf{Proof}}:   Using Proposition \ref{extremePoints}, one selects the two extreme points. Let $a=\min E$ and  $b=\max E$, $a$ and $b$ are computed in $O(n)$ time.
With $\alpha =1$, adding any subset of size $p-2$ of distinct elements of $E\setminus \{a,b\}$, we will have the same MSN dispersion of $b-a>0$ because of the distance additivity in 1D, which is trivially optimal.$\hfill\square$

\section{ $p$-dispersion-MSN is polynomially solvable  in a 2D PF}

Lemma \ref{orderDist} implies Bellman equations for  $p$-dispersion-MSN: % in Proposition \ref{bellmanMSN}, which is proven in Appendix A:
% problem in a 2D PF, which is the key ingredient to design a DP algorithm:

\begin{prop}%[Bellman equations for $p$-dispersion-MSN]
\label{bellmanMSN}
{  Let $E=\{x_i\}_{i \in [\![1,n]\!]}$ be a re-indexed 2D PF.}
Defining $C_{k,i}^{MSN}$ as the optimal cost of  $k$-dispersion-MSN  among the  points  re-indexed in $[\![1,i]\!]$ for all $k \in [\![2,p]\!]$ and  $i \in [\![k,n]\!]$,
we have:
\begin{equation}\label{initMSN}
 \forall i \in [\![1,n]\!], \:\:\: C^{MSN}_{2,i} =  d_{1,i}
 \end{equation}
\begin{equation}\label{inducFormMSN}
  \forall k \in [\![3,p]\!], \: \forall i \in [\![k,n]\!], \:\:\: C^{MSN}_{k,i} =  \max_{j \in [\![k-1,i-1]\!]} \left(C^{MSN}_{k-1,j} + d_{j,i}\right).
\end{equation}
\end{prop}

 \noindent{\textbf{Proof}:} (\ref{initMSN}) is given by Proposition \ref{2Disp}. We suppose $k\geqslant 3$  and prove (\ref{inducFormMSN}).
Let $i \in [\![k,n]\!]$.
Selecting for each $j \in [\![k-1,i-1]\!]$ an optimal solution of $(k-1)$-dispersion-MSN among points indexed in $ [\![1,j]\!]$,
and adding point $i$, it defines a feasible solution for $k$-dispersion-MSN among points indexed in $ [\![1,i]\!]$
with a cost $C^{MSN}_{k-1,j} + d_{j,i}$.
This last cost is lower than the optimal $k$ dispersion cost, thus $C^{MSN}_{k,i} \geqslant  C^{MSN}_{k-1,j} + d_{j,i}$. Therefore
\begin{equation}\label{eqBellmanMSmineq}
 C^{MSN}_{k,i} \geqslant  \max_{j \in [\![k-1,i-1]\!]} (C^{MSN}_{k-1,j} + d_{j,i}).
\end{equation}

Let $ j_1<j_2<\dots<j_{k-1}< j_k$ be indexes %such that $1 = j_1<j_2<\dots<j_{k-1}< j_k = i$
defining an optimal solution of $k$-dispersion-MSN, its cost is $C^{MSN}_{k,i}$.
Because of Proposition \ref{extremePoints}, we can assume that  $j_1 = 1$ and $j_k = i$.
Necessarily, $j_1, j_2,\dots,j_{k-1}$ defines an optimal solution of
 $(k-1)$-dispersion-MSN among points indexed in $ [\![1,j_{k-1}]\!]$. On the contrary, a strictly better solution for $C^{MSN}_{k,i}$ would be constructed adding the index $i$.
We have thus:
$C^{MSN}_{k,i} =  C^{MSN}_{k-1,j_{k-1}} + d_{j_{k-1},i}$.
Combined with (\ref{eqBellmanMSmineq}), it proves : $ C^{MSN}_{k,i} =  \max_{j \in [\![k-1,i-1]\!]} (C^{MSN}_{k-1,j} + d_{j,i})$. $\hfill \square$

\vskip 0.23cm
Algorithm 1 is a  first DP algorithm for $p$-dispersion-MSN based on Proposition \ref{bellmanMSN}.
The first phase computes the matrix of optimal costs $C^{MSN}_{k,i}$ with index $k$ increasing. %, using that $C^{MSN}_{k,i}$ is computed knowing only optimal values $C^{MSN}_{k-1,j}$ that are computed before for $k\geqslant 3$.
 $C^{MSN}_{p,n}$ is the optimal value of MSN $p$-dispersion. Then, backtracking operations in the matrix  $C^{MSN}_{k,i}$  return an optimal solution.
% { The validity and the complexity of Algorithm 1 are given in Proposition \ref{validAlgoMSN}, the proof is written in Appendix A.}

%\subsection{General DP algorithm }
\begin{figure}[ht]
 
  \centering
\begin{tabular}{ l }
\hline
\textbf{Algorithm 1:  $p$-dispersion-MSN in a 2D PF with $p\geqslant3$}\\
\hline
\textbf{Input:}
$n$ points  of a 2D PF, $E=\{x_i\}_{i \in [\![1,n]\!]}$ ; %;\\
an integer $p \in [\![3,n]\!]$.\\
$\phantom{2}$\\
 re-index $E$ following the order of Lemma 1\\
 initialize  matrix $C$ with  $C_{k,i}:=0$  for all $i\in [\![1,n]\!], k\in [\![2;p-1]\!]$\\
% $\phantom{2}$\\
 \textbf{for} $i=1$ to $N-1$:\\ 
\verb!    ! $C_{2,i} := d_{1,i}$\\
 \textbf{end for} \\
% $\phantom{2}$\\
\textbf{for} $k=3$ to $p-1$ :\\
\verb!    ! \textbf{for} $i=k$ to $n-1$:\\
\verb!    ! \verb!    ! $C_{k,i} := \max_{j \in [\![k-1,i-1]\!]} (C_{k-1,j}  +d_{j,i})$\\
\verb!    ! \textbf{end for} \\
\textbf{end for} \\
%$\phantom{2}$\\
 $j := \mbox{argmax}_{j \in [\![p-1,N-1]\!]} (C_{p-1,j} + d_{j,N})$;  $OPT := C_{p-1,j} + d_{j,N}$\\
 initialize $i:=j$ and $\JJ:=\{1,j,N\}$.\\
 \textbf{for} $k=p-1$ to $3$ with increment $k \leftarrow k-1$:\\
\verb!    ! $j := \mbox{argmax}_{j' \in [\![k-1,i-1]\!]} (C_{k-1,j'} + d_{j',i})$\\
\verb!    ! add $j$ in $\JJ$\\
 %\textbf{if} $j>1$ \textbf{then}
\verb!    ! $i:=j$\\
 \textbf{end for} \\
%$\phantom{2}$\\
\textbf{return}  $OPT$ the optimal cost  and the set of selected indexes $\JJ$. \\
\hline
\end{tabular}
\end{figure}

\begin{prop}\label{validAlgoMSN}
{  Let $E=\{x_i\}_{i \in [\![1,n]\!]}$ be a 2D PF},
%Let  $E =\{x_1,\dots, x_n\}$ a subset of $n$ points  of $\RR^2$, %fulfilling hypothesis (\ref{hypoNonDominated})
%such that for all $ i\neq j$, $x_i \mathcal{I} x_j$.
let $p\geqslant 2$.
Algorithm 1 solves  $p$-dispersion-MSN  in $O(pn^2)$ time and $O(pn)$ memory space.
\end{prop}

 \noindent{\textbf{Proof}:} Let  $p\geqslant 3$ and let us prove the validity of Algorithm 1.
 Induction formula (\ref{inducFormMSN}) uses only values $C_{i,j}$ with $j<k$.
 Hence, $C_{k,n}$ is  at the end of each loop in $k$ the optimal value
 of $k$-dispersion-MSN  among the $n$  points of $E$,
 and the optimal cost is given by $C_{p,n}$.
 The remaining operations consist in a  standard backtrack algorithm to return an optimal solution.
 This proves the validity of Algorithm 1 to solve optimally  $p$-dispersion-MSN.

 Let us analyze the complexity.
Re-indexing  $E$ following Lemma \ref{reord} has a time complexity in $O(n\log n)$.
 Computing the line $k=2$ of the DP matrix  has also a time complexity in $O(n)$.
 Computing $\max_{j \in [\![k-1,i-1]\!]} (C_{k-1,j}  +d_{j,i})$  is in $O(i-k)$ and thus in $O(n)$ enumerating all the $i-k$ possibilities.
 It induces  time complexities in $O(pn^2)$ for the construction of the DP matrix, and in $O(pn)$ for the backtracking operations.
  Finally, the  time complexity is given by the construction of the DP matrix, in $O(pn^2)$ time.
 The space complexity is in $O(pn)$, storing the DP matrix $C$. $\hfill\square$

 \vskip 0.23cm
% \noindent{\textbf{Remark}:}\label{remarkOn}

 In Algorithm 1, the DP matrix $C$ is computed line by line, with  index $k$ increasing.
The computation of line $k+1$ requires only line $k$ and $O(1)$ computations of distances. % using $O(1)$ additional memory space.
To compute only the optimal value $C_{p,n}$, it is possible to delete the line $k-1$ once the line $k$ is completed.
Such implementation has  a spatial complexity in  $O(n)$ with at most $2n$ elements in memory.
One may  have only one line in memory, with an ``in-place'' implementation, computing values $C_{k,m}$ for a given $k$ with index $m$ decreasing: in-place $m'$-th values for $m'<m$ are still $C_{k-1,m'}$ that are needed to compute $C_{k,m}$.
Algorithm 1 has a spatial complexity in $O(pn)$ because the backtracking operations use  the full DP matrix. % to compute an optimal solution.
 Algorithm 2  has a $O(n)$ memory space algorithm by adapting techniques that were  used in \cite{choi2021maximizing,advancedDP}.
 Algorithm 2 stores an intermediate value in the middle of the path of an optimal solution, to recover an optimal solution with a recursive divide and conquer strategy which will not be penalizing for the asymptotic time complexity. Such recursion applied to  index $j = \mbox{argmax}_{j' \in [\![p-1,N-1]\!]} (C_{p-1,j'} + d_{j',N})$ is valid, but it would lead to a $O(p^2n^2)$ time complexity to have a space complexity on $O(n)$.

\vskip 0.23cm
Let $p' = \left\lfloor \frac p 2 \right\rfloor $.  We define a DP matrix $H$ with $H_{k,m}$ for $m\in [\![1,N]\!]$ and $k\in \left[\!\left[p'; p\right]\!\right]$
an index in  $[\![1,m]\!]$ such that there is an optimal solution of $k$-dispersion-MSN  among points in $[\![1,m]\!]$
such that the first $p'$  selected points are an optimal solution of  $p'$-dispersion-MSN among points indexed in $[\![1,H_{k,m}]\!]$. We have following induction relations:

\begin{equation}\label{initMSNidxMid}
 {  \forall i \in [\![p',n]\!], \:\:\: H_{p',i} = i}
 \end{equation}
\begin{equation}\label{inducFormMSNidxMid}
 {  \forall k> p', \: \forall i \in [\![k,n]\!], \:\:\: H_{k,i} =  H_{k-1, \mbox{argmax}_{j \in [\![k-1,i-1]\!]} (C^{MSN}_{k-1,j} + d_{j,i})}}
\end{equation}

 \begin{figure}[ht]

  \centering
\begin{tabular}{ l }
\hline
\textbf{{  Algorithm 2:  $p$-dispersion-MSN in a 2D PF using $O(n)$  space}}\\
\hline

\textbf{Input:} \\
- $n$ points  of $\RR^2$, $E=\{x_i\}_{i \in [\![1,n]\!]}$   {a re-indexed 2D PF} ;\\
-  an integer $p$ with $2\leqslant p\leqslant n$.\\
- $a,b \in [\![1,n]\!]$ with $a<b$\\
\textbf{Output:} %\textsc{DivideConquer}($E,a,b,p$) return
 optimal solution and  cost of  $p$-dispersion-MSN in   $\{x_i\}_{i \in [\![a,b]\!]}$ \\
$\phantom{2}$\\
{  \textsc{DivideConquer}($E,a,b,p$)}\\
% re-index $E$ following the order of Proposition 1\\
\verb!    ! \textbf{if} $p=2$ : %\\
%\verb!       !
\textbf{return} $d_{a,b}$, $\{a,b\}$\\
%$\phantom{2}$ \textbf{if} $p=3$ : \\
%\verb!       !  $\phantom{2}$ $c$ := argmax$_{c \in [\![a+1;b-1]\!]} d_{a,c}+ d_{c,b}$\\
%\verb!       !  $\phantom{2}$\textbf{return} $d_{a,c}+ d_{c,b}$, $\{a,c,b\}$\\
% $\phantom{2}$ \textbf{end if} \\
%
\verb!    ! initialize  vector $C$ with  $C_{i}:= d_{a,i}$  for all $i\in [\![a,b]\!]$\\
\verb!    ! initialize  vector $H$ with  $H_{i}:= i$  for all $i\in [\![a,b]\!]$\\
% $\phantom{2}$\\
%  \textbf{for} $i=1$ to $N-1$:\\
%  $\phantom{2}$ $C_{2,i} := d_{1,i}$\\
%  \textbf{end for} \\
% $\phantom{2}$\\
\verb!    ! \textbf{for} $k=3$ to $p-1$ with increment $k\leftarrow k+1$ :\\
\verb!       ! \textbf{for} $i=b$ to $a+k-2$  with increment $i\leftarrow i-1$  :\\
\verb!          !  $j := \mbox{argmax}_{j \in [\![a+k-3,i-1]\!]} (C_{j}  +d_{j,i})$\\
\verb!          !  $C_{i} := C_{j}  +d_{j,i}$\\
\verb!       !  \textbf{if} $k>\left\lfloor \frac p 2 \right\rfloor$ :   $H_{i}:= H_{j}$\\
\verb!       ! \textbf{end for} \\
\verb!    ! \textbf{end for} \\
%$\phantom{2}$\\
\verb!    !  $j := \mbox{argmax}_{j \in [\![p-1,N-1]\!]} (C_{j} + d_{j,b})$;  $OPT := C_{j} + d_{j,b}$ ; %\\
$h := H_{j}$\\
\verb!    ! delete vectors $C$ and $H$\\
\verb!    ! \textbf{if} $p=3$ : %\\
%\verb!       !
\textbf{return} $d_{a,b} +d_{j,b}$, $\{a,j,b\}$\\
\verb!    ! \textbf{else if} $p=4$ : %\\
%\verb!       !
\textbf{return} $d_{a,h} +d_{h,j} + d_{j,b}$, $\{a,h,j,b\}$\\
\verb!    !  \textbf{else : }\\
\verb!       ! $OPT _1,\JJ_1 :=$\textsc{DivideConquer}$\left(E,a,h,\left\lfloor \frac p 2 \right\rfloor \right)$\\
\verb!       !  $OPT _2,\JJ_2 :=$\textsc{DivideConquer}$\left(E,h,j,p-1-\left\lfloor \frac p 2 \right\rfloor \right)$\\
\verb!       ! \textbf{return}  $OPT$,  $\JJ_1 \cup \JJ_2  \cup \{b\}$. \\
\hline
\end{tabular}
\end{figure}

\begin{theorem}%[$k$-means is polynomial for a $2$-dimension Pareto front]
{  Let   $E=\{x_i\}_{i \in [\![1,n]\!]}$ be  a  2D PF.}
Max-Sum-Neighbor $p$-dispersion is  solvable in polynomial time  in the 2D PF $E$.
The cases $p=2,3$ are solvable  in $O(n)$ time
using an $O(1)$ additional memory space.
When $p>3$,  $p$-dispersion-MSN is solvable in $O(pn^2)$ time and $O(n)$ memory space.
\end{theorem}

 \noindent{\textbf{Proof}:} Cases $p=2,3$ are given by 
 Propositions \ref{2Disp} and \ref{trivialCases3}, so that we suppose $p\geqslant 4$ and we consider   Algorithm 2, calling \textsc{DivideConquer}($E,1,n,p$) after a $O(n \log n)$ time re-indexing using Lemma \ref{reord}, which will not influence the final complexity.
 The validity and complexity of the cost computations arez given by Proposition \ref{validAlgoMSN}. %It is the same computation for OPT with a linear space.}
  By induction and using (\ref{initMSNidxMid}), (\ref{inducFormMSNidxMid}), we prove easily that  $H_{m}$ for $m\in [\![1,N]\!]$ is
at the end of loop
 $k\in \left[\!\left[p' , p-1\right]\!\right]$
an index in  $[\![1,m]\!]$ such that there is an optimal solution of $k$-dispersion-MSN  among points in $[\![1,m]\!]$
such that the $\left\lfloor \frac p 2 \right\rfloor$ first selected points are an optimal solution of  $\left\lfloor \frac p 2 \right\rfloor$-dispersion-MSN in $[\![1,H_{m}]\!]$.
For the last iteration $p$, the value of $H_{p,n}$ is $h=H_j$.
The validity of the backtracking operations is given by induction. The terminal cases $p=2$ and $p=3$ are given by Propositions \ref{2Disp} and \ref{trivialCases3}.
The terminal case with $p=4$ is given with the extreme points, $j$ is the third point ($3=4-1$) computed by OPT, and the second point is $h$ ($2 = \left\lfloor \frac 4 2 \right\rfloor$).
  By induction, the optimal solution is concatenated using that  $x_b$ is an optimal selected point, having
an optimal solution  of $p'$-dispersion-MSN calling  \textsc{DivideConquer}$\left(E,a,h,p' \right)$ and it remains to compute a $(p-p'-1)$-dispersion-MSN between $x_h$ and $x_j$.
This proves the validity by induction.

 Space complexity is in $O(n)$ %using $2n$ elements in memory
 with $C$ and $H$ vectors in the first cost computation,
 and thereafter the space usage decreases ($C$ and $H$ are deleted before the recursive calls). Key point is the time complexity.
Let $T(n,p)$ the computation time to compute $p$-dispersion-MSN among $n$ points.
%$T(n,2)$, $T(n,3)$ are $O(1)$ computations, let $\alpha$ be an upper bound.
Using Proposition \ref{validAlgoMSN}, there exists $\beta$ such that  $\beta p n^2$ is an upper bound for  the computation of OPT. Hence, We have following induction relation:
\begin{equation}
 { T(n,p) \leqslant \beta p n^2 + T(H_{p,n},p') +  T(j-H_{p,n},p-p'-1)}
 \end{equation}
 { By induction, we can prove that it exists $\gamma \geqslant 2 \times \beta$ such that for all $n,p$, $T(n,p) \leqslant \gamma p n^2$. It is true for terminal conditions and by induction:}
\begin{equation}
 { T(n,p) \leqslant \beta p n^2 + \gamma p' (H_{p,n})^2 + \gamma (p-p'-1) \times (n-H_{p,n})^2 \leqslant \gamma p n^2}.
 \end{equation}

%
% Propositions \ref{2Disp} implies that
% By induction, calling \textsc{DivideConquer}$\left(E,1,H_{j},\left\lfloor \frac p 2 \right\rfloor \right)$\\
% \verb!       !  $OPT _2,\JJ_2 :=$\textsc{DivideConquer}$\left(E,H_{j},j,p-\left\lfloor \frac p 2 \right\rfloor \right)$
Hence, Algorithm 2  runs in $O(pn^2)$ time using $O(n)$ space. $\hfill\square$

\vskip 0.43cm

  \noindent{\textbf{Remark}}:  { Using Algorithm 2 instead of Algorithm 1, if we have the same asymptotic time complexity, the number of operations (and thus the  CPU time) is approximatively doubled. As mentioned by
   \cite{advancedDP}, this should be used only if memory is missing to use %, which may not be the case with
   Algorithm 1.}

\section{$p$-dispersion-Mm is polynomially solvable in a 2D PF}

Lemma \ref{orderDist} implies Bellman equations for Max-min $p$-dispersion: %  in Proposition \ref{bellman}, which is proven in Appendix A:

\begin{prop}%[Bellman equations]
\label{bellman}
 {Let $E=\{x_i\}_{i \in [\![1,n]\!]}$ be a re-indexed 2D PF.}
Defining $C^{Mm}_{k,i}$ as the optimal cost of Max-min $k$-dispersion among the  points  re-indexed in $[\![1,i]\!]$ for all  $k \in [\![2,p]\!]$ and $i \in [\![k,n]\!]$ ,
we have following  relations:
\begin{equation}\label{initBellman}
 \forall i \in [\![1,n]\!], \:\:\: C_{2,i}^{Mm} =  d_{1,i}
 \end{equation}
\begin{equation}\label{inducForm}\forall k \in [\![3,p]\!],\: 
\forall i \in [\![k,n]\!],  \:\:\: C_{k,i}^{Mm} =  \max_{j \in [\![k-1,i-1]\!]} \min(C_{k-1,j}^{Mm} ,d_{j,i}).
\end{equation}
% Once the $C_{j-1,k-1}$ are computed, the computations of $C_{i,k} =  \max_{j \in [\![k-1,i-1]\!]} \min(C_{k-1,j} ,d_{j,i})$ have a time complexity in $O(\log i)$.
\end{prop}

  \noindent{\textbf{Proof}}: %The proof is similar to the one of Proposition \ref{bellmanMSN}.
  (\ref{initBellman}) is given by Proposition \ref{2Disp}. We suppose $k\geqslant 3$  and prove (\ref{inducForm}).
Let $i \in [\![k,n]\!]$.
Selecting for each $j \in [\![k-1,i-1]\!]$ an optimal solution of $(k-1)$-dispersion-Mm  in $\{x_l\}_{l \in  [\![1,j]\!]}$,
and adding point $i$, it defines a feasible solution for $k$-dispersion-Mm in $\{x_l\}_{l \in  [\![1,i]\!]}$ %among points indexed in $ [\![1,i]\!]$
with a cost $\min (C^{Mm}_{k-1,j} , d_{j,i} )$.
This last cost is lower than the optimal $k$ dispersion cost, thus $C^{Mm}_{k,i} \geqslant \min ( C^{Mm}_{k-1,j} , d_{j,i} )$. Therefore,
%\begin{equation}\label{eqBellmanMmIneq}
$ C^{Mm}_{k,i} \geqslant  \max_{j \in [\![k-1,i-1]\!]} \min\left(C^{Mm}_{k-1,j}, d_{j,i}\right)$.
%\end{equation}

Let $ j_1<j_2<\dots<j_{k-1}< j_k$ be indexes %such that $1 = j_1<j_2<\dots<j_{k-1}< j_k = i$
defining an optimal solution of $k$-dispersion-Mm, its cost is $C^{Mm}_{k,i}$.
Using Proposition \ref{extremePoints}, we can assume that  $j_1 = 1$ and $j_k = i$.
Let $c$ be the $(k-1)$-dispersion-Mm of points indexed by $ j_1<j_2<\dots<j_{k-1}$. We have  $c \leqslant C^{Mm}_{k-1,j_{k-1}}$.
If $c \geqslant d_{j_{k-1},j_k}$, the bottleneck distance is given by points indexed by $j_{k-1},j_k$ and we have  $C^{Mm}_{k,i} = d_{j_{k-1},j_k} =   \min(C^{Mm}_{k-1,j_{k-1}} , d_{j_{k-1},i})$.
Otherwise,  $j_1, j_2,\dots,j_{k-1}$ define an optimal solution of
 $(k-1)$-dispersion-Mm among points indexed in $ [\![1,j_{k-1}]\!]$. On the contrary, a strictly better solution for $C^{Mm}_{k,i}$ would be constructed adding the index $i=j_k$.
We have thus in this case:
$C^{Mm}_{k,i} =  \min (C^{Mm}_{k-1,j_{k-1}} , d_{j_{k-1},i} )$.
This is also true in the case  $c \geqslant d_{j_{k-1},j_k}$,  this is thus always true.
It proves the reverse inequality : $ C^{Mm}_{k,i} \leqslant  \max_{j \in [\![k-1,i-1]\!]} \min (C^{Mm}_{k-1,j} , d_{j,i} )$. $\hfill \square$

 \vskip 0.23cm
 
 As in Algorithm 1, equations (\ref{initBellman}) and (\ref{inducForm}) allow to design a DP algorithm with a complexity in $O(pn^2)$ time and $O(pn)$ space.
  Following developments improve this complexity. Firstly, the time complexity is improved with a logarithmic search in  Algorithm 3:

 \begin{figure}[ht]
 \centering 
\begin{tabular}{ l }
\hline
\textbf{Algorithm 3: Computation of  $\mbox{max}_{j \in [\![k-1,i-1]\!]} \min(C_{k-1,j}^{Mm} ,d_{j,i})$}\\
\hline
%\textbf{input}: indexes $k<i$\\
%\verb!  !  \\
\verb!  ! define $a:=k-1$,  {$b:=i$} \\
\verb!  ! \textbf{while} $b-a\geqslant 2$  \\
\verb!    ! Compute $j= \left \lfloor \frac {a+b} 2 \right \rfloor$\\
\verb!    ! \textbf{if} $C_{k-1,j}^{Mm} - d_{j,i}>0$  \verb! !\textbf{then }\verb! !  $b:=j$ \\
\verb!      ! \textbf{else}  \verb! !  $a:=j$ \\
\verb!  ! \textbf{end while} \\
%\verb!  !\\
\verb!  ! \textbf{return} $\max (\min(C_{k-1,a}^{Mm} ,d_{a,i}) , \min(C_{k-1,b}^{Mm} ,d_{b,i}))$\\

\hline
\end{tabular}
\end{figure}
  
\begin{prop}\label{bellmanComplexity} Let $k \in [\![3,p]\!]$ and $i \in [\![k,n]\!]$.
Algorithm 3 computes  $C_{k,i}^{Mm} =  \max_{j \in [\![k-1,i-1]\!]} \min(C_{k-1,j}^{Mm} ,d_{j,i})$ with a time complexity in $O(\log (i+1-k))$ once the $C_{k-1,j}^{Mm}$ are computed for all $j \in [\![k-1,i-1]\!]$ .
\end{prop}

\noindent{\textbf{Proof}}:
Let $k \in [\![3,p]\!]$ and $i \in [\![k,n]\!]$.
  Lemma \ref{orderDist} ensures that  the application $j \in [\![k-1,i]\!] \mapsto d_{j,i}$ is strictly decreasing.
 The application $j \in [\![k-1,i]\!] \mapsto C_{k-1,j}^{Mm}$ is increasing: any feasible solution of $(k-1)$-dispersion-Mm
 among the $j$ first points, is a feasible solution for $(k-1)$-dispersion-Mm considering the $j+1$ first points, and the optimal value $C_{k-1,j}^{Mm}$ is increasing.
 Hence, $\varphi_{i,k}: j \in [\![k-1, {i}]\!] \mapsto C_{k-1,j}^{Mm}  - d_{j,i}$ is strictly increasing.\\
 Let $\psi_{i,k}: j \in [\![k-1,i]\!] \mapsto \min(C_{k-1,j}^{Mm} ,d_{j,i})$ . Note that
 $\varphi_{i,k}(i) =  C_{k-1,i}^{Mm}>0$.\\
Let $\alpha = \min \{j \in [\![k-1,i]\!], \varphi_{i,k}(j) \geqslant 0\}$.\\
For $j\geqslant \alpha$, $\psi_{i,k}(j)=  d_{j,i}$, and $\psi_{i,k}$ is strictly decreasing for $j\geqslant \alpha$.
For $j<\alpha$, $\psi_{i,k}(j)=  C_{k-1,j}^{Mm}$, and $\psi$ is  increasing for $j< \alpha$.
Hence, $\psi_{i,k}$ reaches a maximum for $j= \alpha$ or $j= \alpha -1$.
%Similarly, any index $\beta \in [\![k-1,i-1]\!]$ such that $\varphi_{i,k}(\beta)=0$ is a minimum for $\psi_{i,k}$.
 The computation of $\alpha$, as the minimal value such that $\varphi_{i,k}(j) \geqslant 0$, %or any value $\beta \in [\![k-1,i-1]\!]$ such that $\varphi_{i,k}(\beta)=0$
can be solved with a dichotomic search presented in Algorithm 3, for a time complexity in $O(\log (i+1-k))$. $\hfill \square$

\vskip 0.23cm

% Proposition \ref{bellmanComplexity} allows to improve
% the time complexity of the DP algorithm for Max-min $p$-dispersion problem.
 To have a linear  space complexity, one can design a recursive DP algorithm as in Algorithm 2. % for  $p$-dispersion-MSN,
For Max-min $p$-dispersion, simple greedy algorithms in Algorithms 4 and 4' are valid as
backtracking procedures: %to compute an optimal solution from the optimal cost. This avoids to double the computation times, as noticed for $p$-dispersion-MSN.
%
% the optimal cost $C_{p,n}^{Mm}$ can be constructed using a memory space in $O(n)$ with in-place operations in a single vector of $n$ elements, deleting the line $k$ of the DP matrix when the line $k+1$ is fully computed.
% The point here is to provide backtracking algorithms which do not  require to have  stored the whole DP matrix $C_{k,i}^{Mm}$, with a complexity in $O(n)$ memory space and $O(pn\log n)$ time.
%Algorithms 4 and 4' compute optimal  solutions of $p$-dispersion-Mm knowing the optimal cost OPT, with greedy strategies:

\begin{figure}[ht]
% \centering
\begin{tabular}{ l }
\hline
\textbf{Algorithm 4: Backtracking algorithm using $O(n)$  space}\\
\hline
% \textbf{input}:\\
% \verb!    !-  $n$ points  of $\RR^2$,$E=\{x_i\}_{i \in [\![1,n]\!]}$  { a re-indexed 2D PF} ;\\
% \verb!    !- $p\in[\![3;n]\!]$;\\
% \verb!    !- $OPT$, the optimal cost of Max-min $p$-dispersion;\\
%\verb!  !  %\textbf{output}: $\II$ an optimal selection of $p$ points of $E$ for  Max-min $p$-dispersion.\\
%\verb!  !\\
\verb!  ! initialize $M := 1$, $m := 1$, $\SS=\{1,n\}$.\\
\verb!  ! \textbf{for} $k=2$ to $p-1$ with increment $k \leftarrow k+1$\\
\verb!    ! $M$ := the smallest index such that $d(x_m,x_M) \geqslant OPT$ \\
\verb!    ! add $M$ to $\SS$\\
\verb!    ! $m := M$\\
\verb!  ! \textbf{end for} \\
\textbf{return} $\SS$ \\
\hline
\end{tabular}

\vskip 0.23cm
\begin{tabular}{ l }
\hline
\textbf{Algorithm 4': Backtracking algorithm using $O(n)$  space}\\
\hline
% \textbf{input}: \\
% \verb!    !-  $n$ points  of $\RR^2$, $E=\{x_i\}_{i \in [\![1,n]\!]}$    {a re-indexed 2D PF} ;\\
% \verb!    !- $p\in[\![3;n]\!]$;\\
% \verb!    !- $OPT$, the optimal cost of Max-min $p$-dispersion;\\
%\verb!     !a vector $v_j =C_{k-1,j}$ for all $j \in [\![1,i-1]\!]$.\\
%\verb!  !  \textbf{output}: $\II$ an optimal selection of $p$ points of $E$ for  Max-min $p$-dispersion.\\
%\verb!  !\\
\verb!  ! initialize $M := n$, $m := n$, $\SS=\{1,n\}$.\\
\verb!  ! \textbf{for} $k=p-1$ to $2$ with increment $k \leftarrow k-1$\\
%\verb!    ! $m := M$\\
%\verb!    ! \textbf{while} $d(m,M) < OPT$ \textbf{do} $m=m-1$ \textbf{end while}\\
\verb!    ! $m$ := the biggest index such that $d(x_m,x_M) \geqslant OPT$ \\
\verb!    ! add $m$ to $\SS$\\
\verb!    ! $M := m$\\
\verb!  ! \textbf{end for} \\
%\verb!  ! add $[\![1,M]\!]$ in $\II$\\
\textbf{return} $\SS$ \\
\hline
\end{tabular}
\vskip 0.23cm
\begin{tabular}{ l }
\textbf{Inputs of Algorithms 4 and 4'}:  % $n$ points  of $\RR^2$,
$E=\{x_i\}_{i \in [\![1,n]\!]}$
{ a re-indexed 2D PF} ;\\
$p\in[\![3;n]\!]$
and $OPT$, the optimal cost of Max-min $p$-dispersion.\\
\textbf{Output}: an optimal solution given by the selected indexes.
\end{tabular}
\end{figure}

\begin{prop}\label{backtrackDP}\label{backtrackOrderLemma}
 {Let  $E=\{x_i\}_{i \in [\![1,n]\!]}$ be a re-indexed 2D PF.}
Let $p \in[\![3;n]\!]$.
Once the  optimal cost of Max-min $p$-dispersion problem is computed,
Algorithms 4 and 4' compute an optimal solution
in $O(p \log n)$ time using $O(p)$ additional memory space.
Furthermore, let $j_1=1,j_2, \dots, j_{p-1},j_{p}=n$ be the indexes of an optimal solution, let $1,i_2, \dots, i_{p-1},n$ (resp $1,i_2', \dots, i_{p-1}',n$) be
the indexes given by Algorithm 4 and  4'. We have:
\begin{equation}%\label{initBellman}
 \forall k \in [\![2,p-1]\!], \;\;\;i_k \leqslant j_k\leqslant i'_k. \end{equation}
In other words,
the indexes given by Algorithm 4 and  4' are lower and upper  bounds
of the indexes of any optimal solution of Max-min $p$-dispersion considering the extreme points $x_1$ and $x_n$.
\end{prop}
\noindent{\textbf{Proof}:} We prove the result for Algorithm 4,  proof for Algorithm 4' is  similar.
Let $j_1=1,j_2, \dots, j_{p-1},j_p=n$ be the indexes of an optimal solution,
let $i_1 = 1,i_2, \dots, i_{p-1},i_p =n$ be the indexes given by Algorithm 4.
Firstly, we prove by induction on $k$ that for all $k \in [\![1,p-1]\!], i_k \leqslant j_k$.
The case $k=1$ is given by $j_1=i_1=1$.
We suppose $k>1$ and that the induction hypothesis is true for $k-1$, i.e. $i_{k-1} \leqslant j_{k-1}$.
The index $i_k$ is the smallest index such that $d(x_{i_k},x_{i_{k-1}}) \geqslant OPT$.
Using Lemma \ref{orderDist} and $i_{k-1} \leqslant j_{k-1}$,
$d(x_{j_k},x_{j_{k-1}}) \leqslant d(x_{j_k},x_{i_{k-1}})$. Having
$i_{k} > j_{k}$ would be in contradiction  with $d(x_{j_k},x_{j_{k-1}}) \geqslant OPT$
and the definition of $i_k$ as the smallest index such that $d(x_{i_k},x_{i_{k-1}}) \geqslant OPT$.
We have also % $d(x_{j_k},x_{j_{k-1}}) \geqslant OPT$.
$i_{k} \leqslant j_{k}$, which terminates the induction proof, indexes $i_k$ are lower bounds of indexes $j_k$.

Let us prove that indexes $i_1 = 1,i_2, \dots, i_{p-1},i_p =n$  define an optimal solution.
By construction  $d(x_{i_k},x_{i_{k-1}}) \geqslant OPT$ for all $k\in [\![1,p-1]\!]$, we have just to prove that 
$d(x_{n},x_{i_{p-1}}) \geqslant OPT$.
Having $i_{p-1} \leqslant j_{p-1} \leqslant  j_p  = n  = i_p$,
 Lemma \ref{orderDist} implies $d(x_{n},x_{i_{p-1}}) \geqslant d(x_{n},x_{j_{p-1}})$.
Optimality implies  $d(x_{n},x_{j_{p-1}}) \geqslant OPT$, and thus  $d(x_{n},x_{i_{p-1}}) \geqslant OPT$.
%Algorithm 3 returns the optimal solution with  minimal indexes,

Let us analyze the complexity.
Algorithm 4 calls at most
$p-2$ times the computation of smallest index $i_k$ such that $d(x_{i_k},x_{i_{k-1}}) \geqslant OPT$,
which can be proceeded with a dichotomic search,
it runs in $O(p \log n)$ time. $\hfill\square$
 \vskip 0.23cm
Using Proposition \ref{backtrackOrderLemma},  Algorithm 5 is a valid DP algorithm for Max-min $p$-dispersion running in $O(n)$ memory space.
Using Algorithms 4 and 4' instead of a divide-and-conquer strategy   avoids to double the computation times , as noticed  for $p$-dispersion-MSN.
Theorem 2 summarizes the complexity results for Max-min $p$-dispersion:
% \begin{figure}[ht]
%  \centering
% \begin{tabular}{ l }
% \hline
% \textbf{Algorithm 5: Max-min $p$-dispersion in a 2D-PF with $p>3$}\\
% \hline
% $\phantom{2}$\\
%
% \textbf{Input:} \\
% - $n$ points  of $\RR^2$, $E=\{x_i\}_{i \in [\![1,n]\!]}$ a 2D PF;\\
% -  an integer $p \in[\![3;n]\!]$.\\
% %\textsc{$p$-dispersion}(E,p)\\
% $\phantom{2}$\\
%  re-index $E$ following the order of Lemma 1\\
% initialize  line $2$ of the matrix $C$ with  $C_{2,i}=0$  for all $i\in [\![1,n]\!]$\\
%  \textbf{for} $i=1$ to $n-1$: \\ %//Construction of the matrix $C$ \\
%  $\phantom{2}$ $C_{2,i} := d_{1,i}$\\
%  \textbf{end for}\\
%  $\phantom{2}$\\
% \textbf{for} $k=3$ to $p$ :\\
%  $\phantom{2}$ initialize  line $k$ of the matrix $C$ with  $C_{k,i}=0$  for all $i\in [\![k;n]\!]$\\
%  $\phantom{2}$ \textbf{for} $i=1$ to $n-1$:\\
% \verb!       ! $C_{k,i} := \displaystyle\max_{j \in [\![k-1,i-1]\!]} \min(C_{k-1,j} ,d_{j,i})$  with Algorithm 2\\
%   $\phantom{2}$ \textbf{end for}\\
%  $\phantom{2}$ delete  line $k-1$ of the matrix $C$\\
%  \textbf{end for}\\
%  $\phantom{2}$\\
% delete  line $p$ of the matrix $C$\\
%  $OPT := C_{p,n}$\\
% $\phantom{2}$\\
% % store in set $\II$ a solution provided by Algorithm 3 or Algorithm 3'\\
% % $\phantom{2}$\\
% \textbf{return}  $OPT$ and a solution of Algorithm 4 (or Algorithm 4')\\
% \hline
% \end{tabular}
% \end{figure}

\begin{figure}[ht]
 \centering
\begin{tabular}{ l }
\hline
\textbf{Algorithm 5: Max-min $p$-dispersion in a 2D-PF with $p\geqslant 3$}\\
\hline
%\verb!   !\\
\textbf{Input:}
$n$ points  of a 2D PF, $E=\{x_i\}_{i \in [\![1,n]\!]}$ ; %;\\
an integer $p \in [\![3,n]\!]$.\\
% \textbf{input:} \\
% - $n$ points  of $\RR^2$, $E=\{x_i\}_{i \in [\![1,n]\!]}$  {a 2D PF};\\
% -  an integer $p \in[\![3;n]\!]$.\\
%\textsc{$p$-dispersion}(E,p)\\
\verb!   !\\
 re-index $E$ following the order of Lemma 1\\
 {initialize  vector $C$ with  $C_{i}= d_{1,i}$  for  $i\in [\![2;n-1]\!]$}\\
\textbf{for} $k=3$ to $p-1$  with increment $k \leftarrow k+1$ :\\
 \verb!   !  {\textbf{for} $i=n-1$ to $k$  with increment $i \leftarrow i-1$}:\\
\verb!       ! $C_{i} := %\displaystyle
\max_{j \in [\![k-1,i-1]\!]} \min(C_{j} ,d_{j,i})$  with Algorithm 3\\
  \verb!   ! \textbf{end for}\\
 \textbf{end for}\\
% \verb!   !\\
 $OPT := \max_{j \in [\![p,n-1]\!]} \min(C_{j} ,d_{j,n})$  with Algorithm 3\\
%\verb!   !\\
\textbf{return}  $OPT$ and a solution of Algorithm 4 (or Algorithm 4')\\
\hline
\end{tabular}
\end{figure}

\begin{theorem}
 {Let $E=\{x_i\}_{i \in [\![1,n]\!]}$ be a 2D PF.}
 Max-min $p$-dispersion  is polynomially solvable to optimality in the 2D PF $E$.
Cases $p=2,3$ are solvable with a complexity in $O(n)$ time
using an additional memory space in  $O(1)$. With $p>3$, Algorithm 5 solves  Max-min $p$-dispersion
with a complexity in $O(pn\log n)$ time and $O(n)$ space.
\end{theorem}
 
  \noindent{\textbf{Proof}:} The cases $p=2,3$ are given by 
 Propositions \ref{2Disp} and \ref{trivialCases3}, so that we suppose $p\geqslant 4$ and we consider Algorithm 5.
 The induction formula (\ref{inducForm}) in Proposition \ref{bellman} uses only values $C_{k-1,j}^{Mm}$  in Algorithm 5.
 At the end of each iteration of the loop in $k$, it holds that  $C_{i}= C_{k,i}^{Mm}$ for all $i\in [\![k;n-1]\!]$.
The optimal cost is given by a last computation with Algorithm 3 to give  $C_{p,n}^{Mm}$.
 The validity of the backtracking procedures is proven in Proposition  \ref{backtrackOrderLemma}.

 Let us analyze the complexity of Algorithm 5. The space complexity is in $O(n)$, storing at most one line of the DP matrix $C^{Mm}$.
Re-indexing  $E$ with Lemma \ref{reord} has a time complexity in $O(n\log n)$.
 Computing the line $k=2$ of the DP matrix  has a time complexity in $O(n)$.
 The other lines are computed in $O(n \log n)$ time with Proposition \ref{bellmanComplexity}, for a total computation of the DP matrix in $O(pn \log n)$ time.
 The backtracking operations run in $O(p\log n)$ time, so that the time complexity is given % by the computation of the DP matrix
 in $O(pn \log n)$ time. $\hfill\square$

\section{ $p$-dispersion-MSm is polynomially solvable in a 2D PF}
  
 To have a DP algorithm for Max-Sum-min $p$-dispersion, more adaptation is needed as the cost computation of adding a new point $i$ depends on the choice of the next selected point $i'>i$. To design a DP algorithm, we do not store partial optimal solution of  Max-Sum-min $k$-dispersion in a subset of points. We define $C_{k,i,i'}^{MSm}$ as the best partial cost of  $k$-dispersion-MSm in $E'=\{x_i\}_{i \in [\![1,i']\!]}$  with $i<i'$ knowing that $i$ is the last selected point before $i'$ and without counting a cost for point $i'$.
% We have following inductions relations between $C_{k,i,i'}^{MSm}$ values:
%Bellman equation can be provided with such definition:

\begin{prop}%[Bellman equations]
\label{bellmanMSm}
 {Let $E=\{x_i\}_{i \in [\![1,n]\!]}$ be a re-indexed 2D PF.}
Let $p \in[\![3,n]\!]$. For all $k \in [\![3,p]\!]$, $i' \in [\![k,n]\!]$ and $i \in [\![k-1,i'-1]\!]$,
Defining $C_{k,i,i'}^{MSm}$ as mentioned before, we have the following induction relations:
\begin{equation}\label{initMSm}
 \forall i' \in [\![3,n]\!], \: \forall i \in [\![2,i'-1]\!], \:\:\: C_{3,i,i'}^{MSm} =  d_{1,i} + \min(d_{1,i},d_{i,i'})
 \end{equation}
 $\forall k \in [\![4,p]\!],
\forall i' \in [\![k,n]\!],  \forall i \in [\![k-1,i'-1]\!],$
\begin{equation}\label{inducForm3}
C_{k,i,i'}^{MSm} =  \max_{ j \in [\![k-2,i-1]\!]} C_{k-1,j,i}^{MSm}  +\min(d_{j,i},d_{i',i}).
\end{equation}
\end{prop}

 \noindent{\textbf{Proof}}:
 { Using Proposition \ref{extremePoints}, $C_{3,i,i'}^{MSm}$ is defined  selecting $1,i,i'$, this makes the partial dispersion removing the $d_{i,i'}$ term as given in (\ref{initMSm}).
 Let $ k \in [\![4,p]\!]$, let $i' \in [\![k,n]\!]$ and let $i \in [\![k-1,i'-1]\!]$.
Selecting for each $j \in [\![k-2,i-1]\!]$ an optimal partial solution of  $(k-1)$-dispersion-MSm among points indexed in $ [\![1,i]\!]$ with $j$ as last selected point before $i$,
and adding point $i$, it makes a feasible solution for  $(k-1)$-dispersion-MSm among points indexed in $ [\![1,i]\!]$
with a partial cost $C_{k-1,j,i}^{MSm}  +\min(d_{j,i},d_{i',i})$.
This last cost is lower than the optimum  $ C_{k,i,i'}^{MSm} \geqslant C_{k-1,j,i}^{MSm}  +\min(d_{j,i},d_{i',i})$. Therefore,
\begin{equation}\label{eqBellmanMSmineqlt}
C_{k,i,i'}^{MSm} \geqslant   \max_{ j \in [\![k-2,i-1]\!]} C_{k-1,j,i}^{MSm}  +\min(d_{j,i},d_{i',i}).
\end{equation}
Let $j_1,\dots,j_k$ be indexes such that $1 \leqslant j_1<j_2<\dots<j_{k-1}=i< j_k = i'$
defining an optimal partial solution of  $k$-dispersion-MSm in $ [\![1,i']\!]$ with $i$ as last selected point before $i'$, its cost is $C_{k,i,i'}^{MSm}$.
Necessarily, $j_1, j_2,\dots,j_{k-1}$ defines an optimal partial solution of
$(k-1)$-dispersion-MSm among points indexed in $ [\![1,i]\!]$ with $j_{k-2}$ as last selected point before $i$. On the contrary, a strictly better solution for $C_{k,i,i'}^{MSm}$ would be constructed adding the index $i'$.
We have thus:
$C_{k,i,i'}^{MSm} =  C_{k-1,j_{k-2},i}^{MSm}  +\min(d_{j_{k-2},i},d_{i',i})$.
Combined with (\ref{eqBellmanMSmineqlt}), it proves : $ C^{MSm}_{k,i} =  \max_{j \in [\![k-2,i-1]\!]} C_{k-1,j,i}^{MSm}  +\min(d_{j,i},d_{i',i})$. $\hfill \square$}

\vskip 0.3cm

 Bellman equations of  Proposition \ref{bellmanMSm} allow to solve $p$-dispersion-MSm in a 2D PF with a DP algorithm detailed in Algorithm 6.  {Once DP matrix  $C_{k,i,i'}^{MSm}$ is computed, the optimal value of the complete  Max-Sum-min $p$-dispersion is the best value $C_{p,j,n} +d_{j,n}$ for $j<n$}.

 \begin{figure}[ht]
 \centering
\begin{tabular}{ l }
\hline
\textbf{Algorithm 6: Max-Sum-min $p$-dispersion in a 2D-PF with $p>5$}\\
\hline
%\\

\textbf{Input:}  $E=\{x_i\}_{i \in [\![1,n]\!]}$ a 2D PF of size $n$ ;  an integer  $p  \in [\![6,n]\!]$.\\
 $\phantom{2}$\\
%\textsc{Max-Sum-min $p$-dispersion}(E,p)\\
 re-index $E$ following the order of Lemma \ref{reord}\\
 initialize  matrix $C$ with  $C_{k,i,i'}:= 0$  for all  $k \in [\![2,p-1]\!]$, $(i,i') \in [\![1,n]\!]^2$ \\
 \textbf{for} $i'=3$ to $n$\\ %//Construction of the matrix $C$ \\
\verb!   !  \textbf{for} $i=2$ to $i'-1$\\
\verb!      !   $C_{3,i,i'} :=  d_{1,i} +   \min(d_{1,i},d_{i,i'})$\\
\verb!   !  \textbf{end for} \\
 \textbf{end for} \\
\textbf{for} $k=4$ to $p$ \\
\verb!   ! \textbf{for} $i'=k$ to $n$\\ %//Construction of the matrix $C$ \\
\verb!      ! \textbf{for} $i=k-1$ to $i'-1$\\
%
%  \textbf{for} $i=1$ to $n$\\ %//Construction of the matrix $C$ \\
%  $\phantom{2}$ \textbf{for} $i'=1$ to $i-1$\\
%  % // case $k=1$ treated separately\\
%  $\phantom{2}$  $C_{3,i,i'}^{MSm} :=  d_{i,i'} +  d_{1,i'} + \min(d_{1,i'},d_{i',i})$\\
% $\phantom{2}$ \textbf{for} $k=4$ to $p$ \\
\verb!          !   {$C_{k,i,i'} :=  \max_{ j \in [\![k-2,i-1]\!]} C_{k-1,j,i}  +\min(d_{j,i},d_{i,i'})$}\\
\verb!      ! \textbf{end for} \\
\verb!   !  \textbf{end for} \\
 \textbf{end for} \\
  $j := \mbox{argmax}_{j' \in [\![p-2,n-1]\!]} (C_{p,j',n} +d_{j',n})$\\
  $OPT :=  C_{p,j,n}+ d_{j,n}$\\
 initialize $i':=n$, $i:=j$ and $\SS:=\{1,j,n\}$.\\
 \textbf{for} $k=p-1$ to $3$ with increment $k \leftarrow k-1$\\
\verb!   ! compute $j:= $arg$\max_{ j' \in [\![k-2,i-1]\!]} C_{k-1,j',i}  +\min(d_{j',i},d_{i',i})$\\
\verb!   ! add $j$ in $\SS$; \;%\\
 %\textbf{if} $j>1$ \textbf{then}
%\verb!   !
$i':=i$; \;%\\
%\verb!   !
$i:=j$\\
 \textbf{end for} \\
  %add $i'$ in $\SS$\\
\textbf{return}  $OPT$ the optimal cost  and the set of selected indexes $\SS$. \\
\hline
\end{tabular}
\end{figure}

\begin{theorem}%[$k$-means is polynomial for a $2$-dimension Pareto front]
 {Let $E=\{x_i\}_{i \in [\![1,n]\!]}$ be a 2D PF.}
Max-Sum-min $p$-dispersion is polynomially solvable to optimality  in the 2D PF $E$.
Using an additional memory space in  $O(1)$,
cases $p=2,3$ are solvable  in $O(n)$ time,
 case $p=4$ (resp. $p=5$) are solvable  in $O(n^2)$ (resp. $O(n^3)$) time.
  Algorithm 6 % (and  respectively Algorithm 7) solve
  solves  the cases $p>5$ % of Max-Sum-min $p$-dispersion
with a complexity in $O(pn^3)$ time and $O(pn^2)$ %(respectively $O(n^2)$ )
space.

\end{theorem}
 
   \noindent{\textbf{Proof}:} The cases $p=2,3,4,5$ are given by 
 Propositions \ref{2Disp}, \ref{trivialCases3} and \ref{trivialCase4}, so that we suppose $p\geqslant 6$ and we consider Algorithm 6.
 The proof of the validity of Algorithm 6 to compute the optimal value and an solution for Max-Sum-min $p$-dispersion
 is similar to Theorems 1 and 2.
 Algorithm 6 computes optimal values of $C_{k,i,i'}^{MSm}$ with $k$ increasing requiring only $C_{k-1,i,i'}^{MSm}$ values.
 By induction, it proves that for all $k$, $C_{k,i,i'}^{MSm}$ has the  optimal value at the end of the loop $k$.
 {The optimal value of Max-Sum-min $p$-dispersion in $E$ is then given by $\max_{j \in [\![p-1,N-1]\!]} (C_{p,j,n} +d_{j,n})$}.
 The remaining operations define a standard backtrack algorithm.
 This proves the validity of Algorithm 6.

 Let us analyze the complexity. The space complexity is in $O(pn^2)$, storing the DP matrix $C$.
 The time complexity is given by the construction of the matrix $C_{k,i,i'}^{MSm}$, ie $O(pn^2)$ operations running in $O(n)$ time
 with  naive enumerations to compute $C_{k,i,i'} =  \max_{ j \in [\![k-2,i-1]\!]} C_{k-1,j,i}  +\min(d_{j,i},d_{i',i})$.
 Algorithm 6 has thus a time complexity in $O(pn^3)$. $\hfill\square$

\vskip 0.43cm
% A corollary is that these algorithms apply also in 1D, which was not studied before to the best of our knowledge:
 A corollary is that these algorithms also apply in 1D, proving for the first time the polynomial complexity of Max-Sum-min $p$-dispersion in 1D:
%Max-Sum-min $p$-dispersion
% which has not been investigated before to our knowledge:
\begin{coro}[$p$-dispersion-MSm is polynomial in 1D]
 { Let $E=\{x_i\}_{i \in [\![1,n]\!]}$ be a set of $n$ distinct real numbers, let $p \geqslant 2$.
The Max-Sum-min $p$-dispersion problem
is polynomially solvable   in  $E$.
The cases $p=2,3$ are solvable with a complexity in $O(n)$ time
using an additional memory space in  $O(1)$ .
The cases $p=4$ (resp $p=5$) are solvable with a complexity in $O(n^2)$ (resp $O(n^3)$) time
using an additional memory space in  $O(1)$.
The cases $p>5$ are solvable in $O(pn^3)$ time and $O(pn^2)$ memory space.}
\end{coro}

\noindent{\textbf{Proof}:} By applying Algorithm 6 and Theorem 3  for $E'=\{(0,x_i)\}_{i \in [\![1,n]\!]}=\{(0,x_1), \dots, (0,x_n)\}$, it would give the result with a valid algorithm.
However, this degenerate case of PF is not considered by our assumptions. Therefore, we use an alternative definition of $E'$ to conform to the assumptions of this article. Using the  Euclidean distance and denoting  $M= \max_{j\in [\![1,n]\!]}x_j$, we  consider the affine 2D PF:
\begin{equation}
E'=\left\{\left( \frac {M-x_i} {\sqrt{2}} ,\frac {x_i} {\sqrt{2}} \right)\right\}_{i \in [\![1,n]\!]} .
\end{equation}
With this definition, we  still have:
\begin{equation}
  \forall (i,j) \in [\![1,n]\!]^2, \;\;\;d\left(\left(\frac {M-x_i} {\sqrt{2}},\frac {x_i} {\sqrt{2}} \right),\left(\frac {M-x_j} {\sqrt{2}},\frac {x_j} {\sqrt{2}}\right)\right) = |x_i - x_j|
\end{equation}
so that $p$-dispersion-MSm in $E$ is the same problem than considering $p$-dispersion-MSm in $E'$. $\hfill\square$

\vskip 0.3cm
 Algorithm 6  uses a memory in $O(pn^2)$.  To compute only the optimal cost, a  space complexity in $O(n^2)$ is obtained deleting elements $C_{k-1,j',i}$ when all the elements  $C_{k,i,i'}$ are computed.
 As for Algorithm 2, it is possible to decrease this space complexity in $O(n^2)$, using recursion and alternative backtrack operations.
Such algorithm is presented in Appendix.
%\subsection{Complexity result}

%   \subsection{max sum min neighbor distances}
% 
%   $O(pn^2)$
%  
%  ordre hierarchique $p$-dispersion puis max sum min
%  
%    \subsection{max sum min}
% 
%    $O(pn^3)$ time   $O(pn^2)$ space
%    
% 
%  \section{Computational results}
% 
%  TODO
%  
%  experiences min globaux à compter, 
%  
%  tables comparatives qualité optimums sur les autres critères
%  
%  combien 
%  
%  temps de calculs et parallélisation

 \section{Discussions}
 
  In this section, we discuss  some theoretical insights and practical applications of Theorems 1, 2, 3 and Algorithms 1, 2, 5 and 6.

 \subsection{Equivalent solutions and hierarchic $p$-dispersion}\label{polishProcedure}
 
 Proposition \ref{backtrackDP} gives tight bounds for the indexes of optimal solution    of Max-min $p$-dispersion in a 2D PF.
 Many optimal solutions may exist for $p$-dispersion-Mm.
 Having an optimal solution, one can identify the bottleneck distance   and  rearrange other selected points without changing  Max-min dispersion.
 %Such situations occur when p is large  and the points  are not regularly distributed in the 2D PF.
The solutions of Algorithms 4 and 4' are very unbalanced, leading to the largest values for the last calculated distances.
 For a practical application, it is natural to wish  well-balanced solutions.

 In order to achieve this,  a bi-objective hierarchic optimization can  rank  optimal solutions of  $p$-dispersion-Mm with  dispersion-MSN. Bellman equations of Propositions   \ref{bellmanMSN} and \ref{bellman} can be extended. Indeed, we can define DP matrix pairs $(C_{k,i}^{'Mm},C_{k,i}^{'MSN})$
denoting the optimal costs with the lexicographic order, optimizing firstly  $k$-dispersion-Mm, and then  $k$-dispersion-MSN among points $\{x_j\}_{j \in [\![1,i]\!]}$ indexed in $[\![1,i]\!]$. We have $C_{k,i}^{'Mm}=C_{k,i}^{Mm}$. % are unchanged, the same as in Algorithm 5.
%the costs $C_{k,i}^{'MSN}$ are computed among the solutions giving the cost $C_{k,i}^{'Mm}$.
Such DP matrices are constructed in $O(pn^2)$, enumerating for each computation $i,k$  possible costs with an intermediate index $j$, and sorting the best current value with  lexicographic order.  Backtracking as in Algorithm 1, the hierarchic optimization is also running in $O(pn^2)$ time and $O(pn)$ space.
%One may also speed-up such cost computations reusing Algorithm 2 and Proposition \ref{bellmanComplexity}, and showing that the optimal and equivalent solutions leading to $C_{k,i}^{Mm}$ are in a plateau, ending at $\alpha$ or $\alpha - 1$. %This speeds up the computations, without proving theoretically improvements in the complexity.

% Algorithms 3 and 3'  reduce the space complexity, but the optimal solutions given may be improved for the practical application.
 Another way to have well-balanced solutions is to design a  polishing heuristic starting from the  solutions given by  Algorithms 4 or 4'. One may use a local polishing procedure for a better balance   considering 3-dispersion-Mm optimizations for consecutive points. Each 3-dispersion-Mm computation is running in $O(\log n)$ using Proposition \ref{bellmanComplexity} as points are already sorted.
Computing the optimal consecutive  Max-min 3-dispersions runs in  $O(p\log n)$, so that even with $O(n)$ such iterations, the total complexity remains in   $O(pn\log n)$ time and $O(n)$ space.

 \subsection{Implementation and parallelization issues}
 
% 
%  We note firstly that implementation may be generic  Mm and MSN $p$-dispersion problems,
%  dealing with DP matrices in indexes $i,k$, with the naive enumeration to compute each coefficient $C_{k,i}$. It is also helpful for the bi-objective hierarchic optimization discussed in Section 8.2.
%  For the specificities of Max-min $p$-dispersion problems discussed in section 6, the dichotomic algorithm and the dynamic management of memory can also be coded with small adaptations.
% %Such factorized implementation will be furnished soon with a C++ code, with an OpenMP parallelization as discussed below.

Time complexity in $O(pn \log n)$ or $O(pn^2)$ can be satisfactory for large scale  $p$-dispersion computations. For a practical speed-up,
Algorithms 1, 2, 5 and 6 have useful properties for efficient implementation and  parallelization. % in multi or many-core environments.
%As mentioned in Theorems 1, 2 and 3,
The bottleneck in time complexity is the construction of the DP matrices. %; it is crucial to implement and parallelize carefully this phase.
The backtracking algorithm is essentially sequential, but with a low complexity;  this phase is not crucial for the global efficiency.
The initial sorting algorithm is significant in the computation times only for  Max-min $p$-dispersion with small values of $p$.
Standard parallelization of sorting algorithms applies, even using General Purpose Graphical Processing Units (GPGPU)  \cite{sintorn2008fast}.

The construction  of the DP matrix requires  independent computations to   compute $(k+1)$-dispersion costs from  $k$-dispersion values.
In Algorithm 1 (and for the lexicographic optimization), there are $O(n^2)$ independent operations in the $k$-th loop. There are $O(n^3)$ independent operations in the  loop $k$ of Algorithm 5.
After the specific improvements for Max-min $p$-dispersion with Algorithms 2 and 4, there are $O(n)$ independent operations to compute each value $C_{k,i}^{Mm}$ for $i\in[\![k,n]\!]$  in $O(\log i)$ time.
In all cases, the  parallelization is straightforward in a shared memory environment like OpenMP or in a distributed environment using Message Passign Interface (MPI).  Parallel implementation requires only $p-3$ synchronizations; this is a good property for the practical efficiency.
Applying LPT (Lowest Processing Times) rules from  \cite{graham1969bounds} for load balancing among the operations that can be computed in parallel,
it is better to calculate the most time-consuming operations first, starting from the highest values of $i$ down to the lowest.

DP  Algorithms 1, 2 and 5 are cache-friendly for an efficient implementation.
Indeed, it
requires only $k$-dispersion values to compute $(k+1)$-dispersion costs.
Since these $k$-dispersion previous values are called several times, it is crucial to access them quickly. %to theses values.
Having the $k$-dispersion values in  cache allows such quick access. % instead of accessing from the RAM memory.
 As written in Algorithm 1, one  have to cache two vectors of size $n$ for an implementation keeping the previous and current lines of the DP matrix in cache. %, to manipulate only quick access  memory.
%This is reasonable even for large values of $n$ for modern computers.
With in-place implementations as in Algorithms 2 and 4,
one may cache only one vector of size $n$, which is useful when cache size becomes a limiting factor. % with   two vectors of size $n$.
% For cache reasons, one may hybridize Algorithms 1 and 2, considering an in-place construction of the DP matrix with a line, for instance copying firstly the line $k-1$ to $k$ for an in-place computation of line $k$.
In backtracking operations of Algorithm 1, the elements  are likely no longer cached, inducing more access time. With the lower complexity of backtracking,  this is not a problem.

Finally, we discuss the possibility of massive parallelization under GPGPU.
Exhaustive  enumerations as in  % on running in $O(pn^2)$ time,
$p$-dispersion-MSN (and the lexicographic optimization) is compatible with GPGPU parallelization.
However, this is not the case for the dichotomic search  in  Algorithm 3. % is not compatible with a GPGPU parallelization.
For Max-min $p$-dispersion, one may parallelize the $O(pn \log n)$ time version with OpenMP or the $O(pn^2)$ time   under GPGPU. %Practical efficiency becomes dependent on the characteristic of the GPU hardware.
Finally, note that a line by line computation of the DP matrix  as in Algorithm 2  is  useful for GPGPU parallelization:  memory  in the GPU hardware  may be  a limiting factor.

% \subsection{Speeding up MSN $p$-dispersion}
%  Lastly, we mention that the local search procedures using 3-dispersion
% are also parallelizable, considering independent subset of points, which is also useful for the practical application.

 \subsection{From 2D PF to 3D PF}

In this paper, we  considered  2D PF generated by bi-objective optimization.
We analyze here the possible extension of the results for larger dimensions, for an application to MOO problems with three or more objectives.
3D PF are generated  in many real-world optimization problems, for instance maximizing robustness and stability   and minimizing financial costs of maintenance planning \cite{dupin2020matheuristics}.
 General 2D  $p$-dispersion problems is a sub-case of 3D PF: these are  affine 3D PF.
 As in the proof of Corollary 1, a similar transformation  applies to consider general 2D instances as 3D PFs.
 $\mathcal{N}\mathcal{P}$-hard complexity  for such   cases with Max-min $p$-dispersion  implies that Max-min $p$-dispersion is also $\mathcal{N}\mathcal{P}$-hard for 3D PF.
Unless $\mathcal{P}=\mathcal{N}\mathcal{P}$, there is no hope to generalize DP algorithms with a polynomial complexity to PF in dimension three and more.
The 1/2 approximation factor is valid in  3D (and higher dimensions) PF for Max-min and Max-Sum $p$-dispersion problems, as it holds for any  metric space \cite{tamir1991obnoxious,tamir1998comments}.

Lemmas 1 and 2  are fundamental results to design  DP algorithms and
are highly specific for 2D PF cases, no such total order exists in 3D and larger dimensions.
Generally, this explains why  clustering and selection problems  are polynomial in 2D PF  and $\mathcal{N}\mathcal{P}$-hard in larger dimensions.
%Note that  $p$-dispersion-MSN variant is defined only in  2D PF, using also Lemmas 1 and 2, contrary to $p$-dispersion-MSm,  which is generically defined.
 Proposition 1 eases the calculation of $p$-dispersion problems with the selection of extreme points,
which is crucial for Propositions 2, 3, 4 and 5.
An open question is whether Proposition 1 could be extended to 3 dimensions (and higher dimensions).
Extreme points are generically defined by   different lexicographic optimization with permutations of the considered objectives.
An extension of Proposition 1 would be to analyze if
optimal solutions of $p$-dispersion problems in a PF should necessarily contain such extreme points.
Actually, the answer is negative, as shown by the following counter-example.
For $n=5$ and with 3-dispersion problems, we consider points $(10,0,0)$ ; $(5,5,0)$; $(0,10,0)$; $(9.99,0.01,1)$ and $(0.01,9.99,1)$.
There are here only two extreme points (out of the 3! = 6 lexicographic minimizations): $(10,0,0)$ and $(0,10,0)$,
whereas each 3-dispersion variant has the same and unique optimal solution  $(5,5,0)$; $(9.99,0.01,1)$ and $(0.01,9.99,1)$. %, which contains no extreme points.

Lastly, the possible use of 2D PF DP algorithms as heuristics  is discussed for 3D PF and higher dimensions.
 In general, one can project any PF structure into 1D structures, for example with weighted-sum  scalarization, Principal Component Analysis (PCA) or linear regression.
It allows % %and to table the best solution
to initialize a local search approach with solutions obtained after a projection in 1D, as in \cite{huang2021comparing}. % with several such initial solutions.}
The projection of a 3D PF to a representative 2D PF is difficult in general, affine 3D PF represent any planar instance without any regularity. This perspective would hold only for some specific 3D PFs.
%This may be possible for some specific instances, like the previous counter example which is very close to a 2D PF structure. 
%  In such cases, 1D DP algorithms converge quickly, running in $O(\max\{pn,n \log n\})$  time \cite{ravi1994heuristic,wang1988study}.

\subsection{Having continuous 2D PFs}\label{sec::continuousPF}
 The first hypothesis of our paper was to have a 2D PF of size $n$.
To address complexity results for $p$-dispersion problems (and also for $k$-means, $k$-medoids, k-center variants) in the general and in the 2D PF case, a finite number of points shall be considered. This hypothesis can be reformulated as ``let $n$ points from a 2D PF" to define the problem. In the context of PFs, this finite hypothesis may be in contradiction with the possibility of having infinite PFs in MOO problems.
In continuous PFs, $p$-dispersion problems make sense as continuous optimization problems.
If bounded discrete MOO problems induce finite PFss, this is not the case with MO Linear Programs (MOLP) or MO MILPs \cite{ehrgott2003multiobjective}.
For MOLPs, PFs are given as a connected subset of the frontier of a polyhedron, which is described by a finite number of extreme points of this polyhedron. In 2D, extreme points define the extremity of consecutive segments.
For MO MILPs, PFs may be composed of isolated points and PFs of sub MOLPs.
 %In these cases, corresponding to many real world application based on MILP formulations, an analytic formula is available to define the infinite PF.
Continuous MOO problems may also give an analytic formula of PFs.
To address $p$-dispersion problems in such a context, one may sample regularly the continuous parts of PFs to have a good discrete approximation of the dispersion problem.
By analyzing which maximum value of $n$ induces reasonable computation times depending on the context, this guides the choice of the granularity of such a discretization.

\subsection{Application to $k$-medoids/$k$-means clustering in a 2D PF}

 Exact DP algorithms for $k$-medoids  run
in $O(n^3)$ time \cite{dupin2019medoids}.
Such complexity is a bottleneck for the practical applications  with large values of $n$.
Classical heuristics for $k$-means/$k$-medoids problems may be used in such cases \cite{Celebi,schubert2018faster}.
Such heuristics have no guarantee of optimality, and depend heavily on initialization strategies \cite{Celebi}.
One may initialize  local search with   $p$-dispersion  solutions  to have a quick initialization procedure.
Several such initialization strategies are possible.
Firstly, one can initialize the $k$ centroids using $k$-dispersion.
Secondly, one can solve $2k+1$-dispersion, and select the $k$ intermediate points following order of Lemma \ref{reord}.
Thirdly,  one can solve $3k$-dispersion, and select the $3k'+1$ intermediate points for $k' \in [\![0;k-1]\!]$.
For these strategies, one may use optimal DP algorithms or local search iterations of  3-dispersion as in Section \ref{polishProcedure}.
Based on the preprint version of this paper, numerical results were provided for randomly generated 2D PFs with $n\leqslant 5000$. Good results and very quick computation times are obtained with such heuristics, whereas  computing optimal solution with exact DP is time consuming for  $n= 5000$ \cite{huang2021comparing}.
Having many different initialization procedures  is useful for the final quality of solutions, implementing several local searches with different initialization strategies in parallel environments, as in \cite{dupin2018parallel}.
This is also an interest of the numerical results presented in \cite{huang2021comparing}.

\subsection{Applications to MOO meta-heuristics}% and Skyline operator}

% To select and visualize five representative points in a 2D PF for a human decision maker, enumeration algorithm induced by Proposition 6 may be sufficient.
% Complexity results of sections 5, 6, and 7 are useful
% With larger values of $p$,  $O(n^{p-2})$ time computations are intractable.

%Note that a 2D Skyline operator is  also a 2D PF \cite{borzsony2001skyline,cabello2020faster}.
% Selecting k dispersed points in a skyline is a topic of interest  \cite{lin2007selecting,valkanas2013skydiver,cabello2020faster}.
% However,  application hypotheses are different with bi-objective optimization.
% In database requests, returned points are not necessarily in the skyline.
% Dispersion problem has a different nature for skyline operators, one have a large number of points $n$, a skyline of size $h \leqslant n$ that is also large
% where k dispersed point are wished \cite{cabello2020faster}.
% In our hypotheses, we have implicitly $n=h$, which applies naturally for exact MOO methods.
% Indeed, scalarization methods compute points that are necessarily in the PF, exact bi-objective optimization methods like epsilon constraints or  two-phase method construct a PF enumerating only non-dominated points  \cite{visee1998two}.

In the case of population MOO meta-heuristics like EAs, selection operators may be called iteratively among the current $n$ non-dominated points to operate cross-overs or mutations (or applying a trajectory local search)  among a restricted number of $p \ll n$ solutions \cite{talbi2009metaheuristics}.
 Randomized selection operators  may be used in that goal \cite{talbi2009metaheuristics}.
Deterministic strategies can also  provide representative and diverse solutions, in addition to or as a replacement for random  selection operators for some iterations.
 In such context, %Evolutionary Algorithms (EA),
the number of points to select $p$ may be large,  regarding the maximal size of the archive, the maximal number of points that would be in memory.
 Selecting such $p$ points algorithms running in $O(pn\log n)$ time are of major interest in such context, it is the case with Max-min $p$-dispersion, continuous $p$-center problems \cite{dupin2021unified}, and hypervolume subset selection  \cite{kuhn2016hypervolume}.
Discrete $p$-center is even faster to compute in $O(n\log n)$ time \cite{cabello2020faster}.
Note that it is not required to have optimal solutions of such problems inside EAs.
Even k-medoids heuristics  may be used as in \cite{huang2021comparing} as long as the calculation time of the heuristic remains fast,
complexity calculations guide for such an heuristic design.
Polishing procedures are of interest to have better balanced solutions as presented in Section \ref{polishProcedure}.
The fastest DP algorithms may be used to initialize local search heuristics for other problems, as in \cite{huang2021comparing}.

Having several types of  deterministic selection operators   is of interest for EAs to diversify the points where mutations or intensification are operated among consecutive iterations.
On the contrary, it is a source of inefficiency to call the operators of mutations and selections always on the same points.
Varying the value of $p$  gives a first solution of this problem for deterministic operators.
Selecting points in a 2D PF, many optimization measures and algorithms are available, and have different properties; another solution is to solve iteratively different selection problems.
When selecting points in a PF, the rule of thumb is using hypervolume measures and variants  \cite{falcon2020indicator,guerreiro2020hypervolume}.
 HSS has useful properties to explain such popularity  \cite{guerreiro2020hypervolume}.
Clustering with $k$-means or $k$-medoids define dense zones in the PF where little intensification is needed \cite{dupin2019medoids}.
%In such cases, $k$-dispersion  may be a useful initialization strategies for standard local search approach, to avoid the time complexity in $O(n^3)$, as discussed in section 8.4.
 Partial $p$-center variants  detect  outliers (isolated points) simultaneously with a rough definition of clusters \cite{dupin2021unified}.
In a 2D PF, isolated points should be preferred to try intensification strategies in such zones, to have better-balanced points along the 2D PF \cite{dupin2021unified}. 
Without the partial extension, $p$-center problems are faster to compute, but the clustering and selection are less relevant; fast DP algorithms for $p$-center problems are useful to design quick heuristics fr other selection or clustering optimization.
HSS and $p$-dispersion measures  find diverse points in the 2D PF, but can lead to very different solutions: HSS would avoid points that are close to local Nadir points (of knee-points) contrary to $p$-dispersion problems.

An open perspective is to design EAs combining randomized selection operators with different deterministic and complementary strategies.

% Note than in a dynamic and iterative context, selecting $k$ points with a same algorithm in a partial PF that has only marginally changed is useful
% to store previous intermediate computations for DP algorithms, but this has few interest  processing intensification operators always on the same points.

\section{Conclusion and perspectives}

The properties of the four standard $p$ dispersion problems have been examined in a 2D PF using Euclidean, Minkowski or Chebyshev distances. A novel variant, namely Max-Sum-Neighbor $p$-dispersion, is defined specifically for 2D PF.
  Cases $p=2$ and $p=3$ induce a  complexity in $O(n)$  time.
  Cases $p=4$ and $p=5$ have respectively time  complexities in $O(n^2)$ and $O(n^3)$ time.
  Such results are useful to select a small number of representative solutions for decision makers.
%This offers another perspective,  comparing the different variants of $p$-dispersion in large 2D PF with small values $p$ as in \cite{erkut1991comparison}.

Three variants are proven solvable in polynomial time in a 2D PF, with the design of DP algorithms.
  Standard Max-min $p$-dispersion problem is  solvable  in a 2D PF in $O(pn\log n)$  time  and using $O(n)$ memory  space.
 Max-Sum-Neighbor $p$-dispersion  is  solvable  in a 2D PF   in $O(pn^2)$  time  and $O(n)$  space.
 A lexicographic optimization considering  Max-min  and Max-Sum-Neighbor $p$-dispersion is also solvable in $O(pn^2)$  time using $O(pn)$  space.
 Max-Sum-min $p$-dispersion  is solvable  in a 2D PF in $O(pn^3)$  time.
 This last result and the DP algorithm proves also that  Max-Sum-min $p$-dispersion  is polynomially solvable in 1D, which was never studied before.
 Perspectives may be to improve this complexity result using specificity of 1D instances.
Considering Max-Sum $p$-dispersion, the $\mathcal{N}\mathcal{P}$-hardness of 2D PF and also 2D sub-cases are still open questions.
 %, this result is of theoretical interest more than a practical interest in comparison with the two other variants.

 These results are not only of a theoretical interest, but raise also practical perspectives.% compared to  $\mathcal{N}\mathcal{P}$-hardness of 2D or metric spaces sub-cases,
% or  polynomial complexities with 1D or tree sub-cases.
  Complexity  for Max-min  and Max-Sum-Neighbor $p$-dispersion allows a straightforward application
 for large 2D PF. Furthermore,  DP algorithms have useful  properties for an efficient implementation,
 including efficient parallelization in a multi or many-core environment.
It allows an application inside MOO population meta-heuristics to archive partial PF at each iteration.
In this context, $p$-dispersion DP algorithms  may be used also to initialize $k$-medoids clustering in a 2D PF.
%Properties of DP algorithms allow also to choose a relevant value of $p$ in a tolerance range.
 The extension of  results to higher dimensions was discussed.
3D PF cases are $\mathcal{N}\mathcal{P}$-hard and approximation algorithms with factor 1/2 are available.
Perspectives are only to design quick heuristics for PF in such dimensions.
% The $\mathcal{N}\mathcal{P}$-hardness of Max-Sum $p$-dispersion in the planar case is an open question. This work would tend us to conjecture that the 2D PF case is $\mathcal{N}\mathcal{P}$-hard.
%Max-min $p$-dispersion is $\mathcal{N}\mathcal{P}$ hard in the planar case, it is thus also the case for three dimensional PF, there is no hope to extend the results of this paper in dimension 3. The perspective is only to design quick heuristics for PF in dimensions higher than 3.
 Lastly, the results of this paper may be extended to implicit versions of the problem related to Skyline Operator and MOO  applications.
% If a Skyline Operator is similar to a PF, the Skyline is not known and computed by advance.
% The extended problem would consider simultaneously  exploration of points that may be out of the Skyline with the computation of the p most diverse points in the Skyline.

  \subsection*{Acknowledgements}
  The author wishes to thank Professor Arie Tamir for his useful suggestions based on the ArXiv preprint.
   The author gratefully thanks the anonymous reviewers for their helpful comments that allowed to improve the paper very significantly.

\footnotesize
 \bibliographystyle{abbrv}
  \bibliography{biblioCluster.bib}
\normalsize

% \newpage
  \section*{Appendix : DP  with quadratic memory space for $p$-dispersion-MSm}

Similarly to Algorithm 2,
memory space of DP in  Algorithm 6  can be reduced from $O(pn^2)$ to  $O(n^2)$. %, we use similar ideas to Algorithm 2.
  Let $p' = \left\lfloor \frac p 2 \right\rfloor$.
We define  DP matrices $H, H'$ with $H_{k,i,i'}$ and $H'_{k,i,i'}$ defined  for $k\in \left[\!\left[p'; p\right]\!\right]$, $i' \in [\![k,n]\!]$ and $i \in [\![k-1,i'-1]\!]$
 such that there is an optimal solution of $k$-dispersion-MSm  among points in $[\![1,i']\!]$ and selecting $i$ just before $i'$
such that the $p'$ first selected points are an optimal solution of  $p'$-dispersion-MSN in $[\![1,H'_{k,i,i'}]\!]$ selecting  $H_{k,i,i'}$ as $p'$-th point . Such definition induces following induction relations:

\begin{equation}\label{initMSmidxMid1}
 \forall i' \in [\![p',n]\!], \forall i \in [\![p'-1,i'-1]\!]\:\:\: H_{p',i,i'} = i
 \end{equation}
\begin{equation}\label{initMSmidxMid2}
 \forall i' \in [\![p',n]\!], \forall i \in [\![p'-1,i'-1]\!]\:\:\: H_{p',i,i'}' = i'
 \end{equation}
 Denoting for $k>p',  i' \in [\![p',n]\!],  i \in [\![p'-1,i'-1]\!]$:

 $$j_{k,i,i'} =  \mbox{argmax}_{ j' \in [\![k-2,i-1]\!]} C_{k-1,j',i}  +\min(d_{j',i},d_{i',i}):$$ %we have following propagation:

\begin{equation}\label{inducFormMSnidxMid1}
 \forall k> \left\lfloor \frac p 2 \right\rfloor, \forall i' \in [\![p',n]\!], \forall i \in [\![p'-1,i'-1]\!] \:\:\: H_{k,i,i'} =  H_{k-1,j_{k,i,i'},i}
\end{equation}

\begin{equation}\label{inducFormMSnidxMid2}
 \forall k> \left\lfloor \frac p 2 \right\rfloor, \forall i' \in [\![p',n]\!], \forall i \in [\![p'-1,i'-1]\!] \:\:\: H_{k,i,i'}' =  H_{k-1,j_{k,i,i'},i}'
\end{equation}

 { Another difficulty is for the recursion that computation of costs with extreme
depends on the previous and next points in the partial solution.
\textsc{DivideConquer}($E,prev, a,b,next,p$) compute the optimal $p$-dispersion-MSm
 in   $\{x_i\}_{i \in [\![a,b]\!]}$, knowing that the point before $a$ is $prev$, and the point
 after $b$ is $next$. Having $a=1$ or $b=n$ disregard these values, that are set to $prev=-1$
and  $next=n+1$. To ease some reading, we take convention $d_{-1,i} = + \infty$ and
 $d_{i,n+1} = + \infty$.
 The optimal solution of  $p$-dispersion-MSm is then obtained calling \textsc{DivideConquer}($E,-1,1,n,n+1,p$). }

 \begin{figure}[ht]

  \centering
  \small
\begin{tabular}{ l }
\hline

 { \textbf{Algorithm 7:  $p$-dispersion-MSm in a 2D PF using $O(n^2)$  space}}\\
\hline
% \textbf{Input:} \\
% - $n$ points  of $\RR^2$, $E=\{x_i\}_{i \in [\![1,n]\!]}$  a re-indexed 2D PF ;\\
% -  an integer $p$ with $2\leqslant p\leqslant n$.\\
% - $a,b \in [\![1,n]\!]$, $prev,next\in [\![-1;n+1]\!]$ with $prev <a<b<next$\\
% %\textbf{Output:} %\textsc{DivideConquer}($E,a,b,p$) return
%  %optimal solution and  cost of  $p$-dispersion-MSN in   $\{x_i\}_{i \in [\![a,b]\!]}$ kn \\
% $\phantom{2}$\\
\textsc{DivideConquer}($E,prev, a,b,next,p$)\\
% re-index $E$ following the order of Proposition 1\\
\verb!   ! \textbf{if} $p=2$\\
\verb!     !  \textbf{return} $( \min(d_{prev,a},d_{j,a}) + \min(d_{j,a},d_{j,b}) + \min(d_{j,b},d_{b,next})$, $\{a,b\}$\\
\verb!   ! \textbf{if} $p=3$ : \\
\verb!      !  $j' :=  $arg$\max_{j\in  [\![a+1,b-1]\!]} \left( \min(d_{prev,a},d_{j,a}) + \min(d_{j,a},d_{j,b}) + \min(d_{j,b},d_{b,next}) \right)$\\
\verb!      ! \textbf{return} $ \min(d_{prev,a},d_{j',a}) + \min(d_{j',a},d_{j',b}) + \min(d_{j',b},d_{b,next})$, $\{a,j',b\}$\\
%$\phantom{2}$ \textbf{if} $p=3$ : \\
%\verb!       !  $\phantom{2}$ $c$ := argmax$_{c \in [\![a+1;b-1]\!]} d_{a,c}+ d_{c,b}$\\
%\verb!       !  $\phantom{2}$\textbf{return} $d_{a,c}+ d_{c,b}$, $\{a,c,b\}$\\

% $\phantom{2}$ \textbf{end if} \\
%
\verb!   ! initialize  matrices $H,H'$ with  $H_{i,i'}:= i$, $H_{i,i'}':= i'$   for all $i\in [\![a;b-1]\!], i'\in [\![i+1;b]\!]$\\
\verb!   ! initialize   $C$ with  $C_{i,i'}:=   \min(d_{prev,a} ,d_{a,i}) +   \min(d_{a,i},d_{i,i'}) $  for  $i\in [\![a;b-1]\!], i'\in [\![i+1;b]\!] $\\
% $\phantom{2}$\\
%  \textbf{for} $i=1$ to $N-1$:\\
%  $\phantom{2}$ $C_{2,i} := d_{1,i}$\\
%  \textbf{end for} \\
% $\phantom{2}$\\

\verb!   !\textbf{for} $k=3$ to $p-1$ \\
\verb!      ! \textbf{for} $i'=b$ to $a+k$\\ %//Construction of the matrix $C$ \\
\verb!         ! \textbf{for} $i=a+k-1$ to $i'-1$\\
%
%  \textbf{for} $i=1$ to $n$\\ %//Construction of the matrix $C$ \\
%  $\phantom{2}$ \textbf{for} $i'=1$ to $i-1$\\
%  % // case $k=1$ treated separately\\
%  $\phantom{2}$  $C_{3,i,i'}^{MSm} :=  d_{i,i'} +  d_{1,i'} + \min(d_{1,i'},d_{i',i})$\\
% $\phantom{2}$ \textbf{for} $k=4$ to $p$ \\
\verb!          !  $j' :=  $arg$\max_{ j \in [\![1,i-1]\!]} C_{k-1,j,i}  +\min(d_{j,i},d_{i',i})$\\
\verb!          !  $C_{k,i,i'} :=   C_{k-1,j,i}  +\min(d_{j,i},d_{i',i})$\\
\verb!          !  \textbf{if} $k>\left\lfloor \frac p 2 \right\rfloor$ :   $H_{i,i'}:= H_{j,i}$ ; $H_{i,i'}':= H_{j,i}'$ \\
\verb!          ! \textbf{end for} \\
\verb!      !  \textbf{end for} \\
\verb!   ! \textbf{end for} \\
\verb!   !  $j := \mbox{argmax}_{j' \in [\![p-1,n-1]\!]}$ $C_{p-1,j',b} +\min(d_{j',b},d_{b,next})$  ; \\
\verb!   ! $OPT :=  C_{p,j,b}  +\min(d_{j,b},d_{b,next}) $\\
\verb!   ! $h := H_{j,b}$\\
\verb!   ! $h' := H_{j,b}'$\\
\verb!   ! delete matrices $C,H,H'$\\
\verb!   ! \textbf{ if} $p=4$ : %\\
%\verb!       !
\textbf{return} $OPT$, $\{a,h,j,b\}$\\
\verb!   ! \textbf{ if} $p=5$ : %\\
%\verb!       !
\textbf{return} $OPT$, $\{a,h,h',j,b\}$\\

\verb!    !  \textbf{else : }\\
\verb!       ! $OPT _1,\JJ_1 :=$\textsc{DivideConquer}$\left(E,prev,a,h,h',\left\lfloor \frac p 2 \right\rfloor \right)$\\
\verb!       !  $OPT _2,\JJ_2 :=$\textsc{DivideConquer}$\left(E,h,h',j,b,p-1-\left\lfloor \frac p 2 \right\rfloor \right)$\\
\verb!       ! \textbf{return}  $OPT$,  $\JJ_1 \cup \JJ_2  \cup \{b\}$. \\
\hline
\end{tabular}
\end{figure}

\end{document}